%% file: Main.tex
\documentclass[%
 reprint,
nofootinbib,
 amsmath,amssymb,
 aps,
]{revtex4-1}
\usepackage{graphicx}
\usepackage{dcolumn}
\usepackage{bm}
\usepackage{color}
\usepackage{soul}
\usepackage{hyperref}

\begin{document}

\preprint{APS/123-QED}

\title{Moore-Read state in Half-filled Moir\'e Chern band from three-body Pseudo-potential}

\author{Lu Zhang}
 \email{lzhangdh@connect.ust.hk}
 \affiliation{%
 Department of Physics, Hong Kong University of Science and Technology, Clear Water Bay, Hong Kong, China
}%
 
\author{Xue-Yang Song}%
 \email{songxy@ust.hk}
 \affiliation{Department of Physics, Massachusetts Institute of Technology, Cambridge MA 02139,  USA}

\date{\today}
\begin{abstract}
The moir\'e system provides a tunable platform for exploring exotic phases of materials. This article shows the possible realization of a non-Abelian state characterized by the Moore-Read wavefunction in a half-filled moir\'e Chern band, exemplified by twisted $\rm MoTe_2$. This is achieved by introducing short-range repulsive three-body interaction. Exact diagonalization is employed to examine the spectrum in finite size. The incompressibility of the system, the degeneracy of the ground states, and the number of low-energy states provide compelling evidence to identify the ground state as the Moore-Read state. We further interpolate between the three-body interaction and Coulomb interaction to show a phase transition between the composite Fermi-liquid and the Moore-Read state. Finally, we consider the effect of band mixing and derive the three-body interaction using perturbation theory. By exploring the conditions under which band mixing effects mimic short-range repulsive three-body interaction we provide insights towards realizing non-Abelian phases of matter in the moir\'e system.

\end{abstract}

\maketitle

\textit{Introduction}.--- The fractional quantum Hall effect (FQHE) has attracted widespread interest over the past few decades. A series of Abelian FQHE states have been observed experimentally when the Landau level is occupied at specific fillings $\nu = \frac{p}{2sp+1}\ s,p\in \mathbb{Z}$. Meanwhile, non-Abelian states have been proposed theoretically to be the ground state of the strong Coulomb interaction at even-denominator fillings \cite{MooreNonabelian,XiaoGangNonabelian,XiaoGangPaired,RezayiIncompressible,ReadPaired,HaldaneEntanglement}
, which hold significant practical potential for realizing topological quantum computation\cite{KitaevFault,freedman2003topological,RMPnonabelian}. One well-studied non-Abelian quantum Hall state is the Moore-Read Pfaffian state \cite{MooreNonabelian}, which is formed by $p+ip$ -wave pairing of the composite fermions.
While physicists have explored the Moore-Read state for many years and believed that the ground state of 5/2 filling of Landau level stays in the same universal class as the Moore-Read Pfaffian state\cite{willett1987observation,storni2010fractional,bonderson2006detecting,wojs2010skyrmions}, mysteries and questions still regarding realization of the state remain. The Moore-Read state is not particle-hole symmetric which means it cannot be the ground state of the Coulomb interaction with particle-hole symmetry. Physicists have tried various methods to overcome this issue including the spontaneous particle-hole symmetry breaking and band-mixing in quantum hall system(QHS)\cite{MacDonaldmixing,Rezayimixing,Rezayimixing2,Dassarma2008spontaneous,wang2009particle}, though the nature of the state is still under debate\cite{banerjee2018observation}. 
A more realistic question is if one can go beyond the QHS to realize such a state in other 2D platforms. Advancements in 2d materials result in hallmarks of realizing non-Abelian states, with incompressible state observed at even-denominator filling in graphene systems\cite{zibrov2018even,kang2024observation,Zhang2023TwoDimensional}

Recently, the discovery of the fractional Quantum anomalous Hall effect (FQAH) in the twisted bilayer of transition metal dichalcogenides (TMDs)\cite{JiaqiFQAH} and pentalayer rhombohedral stacked graphene \cite{ZhengguangFractional} might shed light on these questions. Several theoretical and numerical works have confirmed the similarity between the flat Chern band \cite{roy2014band,LiangFuFQAH,AshvinCFL,LiangFuCFL,XiaodiFractional,song2024phase,devakul2021magic} and the lowest Landau level in certain parameter regime. Specifically, there is compelling evidence that the ground state of the half-filled Chern band is a composite Fermi liquid at zero field. 
It is expected that a phase transition from the composite Fermi liquid to the paired state, i.e. the Moore-Read state, might be induced by adding sufficiently large perturbations\cite{bonestell1999singular, metlitski2015cooper}. To date, there is no evidence of the Moore-Read state in the QHS at half-filling. 
However, in the realistic system the energy scale of Coulomb interaction $\frac{e^2}{\epsilon_0,\epsilon_r a_M}$, with $a_M$ the superlattice constant, surpasses the band gap between the valence band and conduction band, which means that the effect of the band mixing is large compared to the electron gas \cite{BoThreebody,YangMix}.Another notable point is the absence of exact particle-hole symmetry at half-filling in the Chern band of twisted bilayer TMDs. Furthermore, the complex details of the single particle wave function can lead to various types of interaction according to the perturbation theory in contrast to the electron gas.
Therefore the complexity of moir\'e materials opens up the possibility for the system to host the Moore-Read state in the Chern insulator.

In this work, we demonstrate the potential existence of the non-Abelian Pfaffian state in twisted $\rm MoTe_2$ and present a condition to realize the state in realistic settings.
We begin by introducing the minimal Hamiltonian for the twisted bilayer $\rm MoTe_2$ by a continuum model with Coulomb interaction and complement the model further with three-body interaction.
We choose the simplest short-range repulsive three-body interaction defined on the triangular lattice to capture the low energy physics of the band mixing effect and break the particle-hole symmetry.
\begin{figure}[h]
    \centering
    \includegraphics[width=0.9\linewidth]{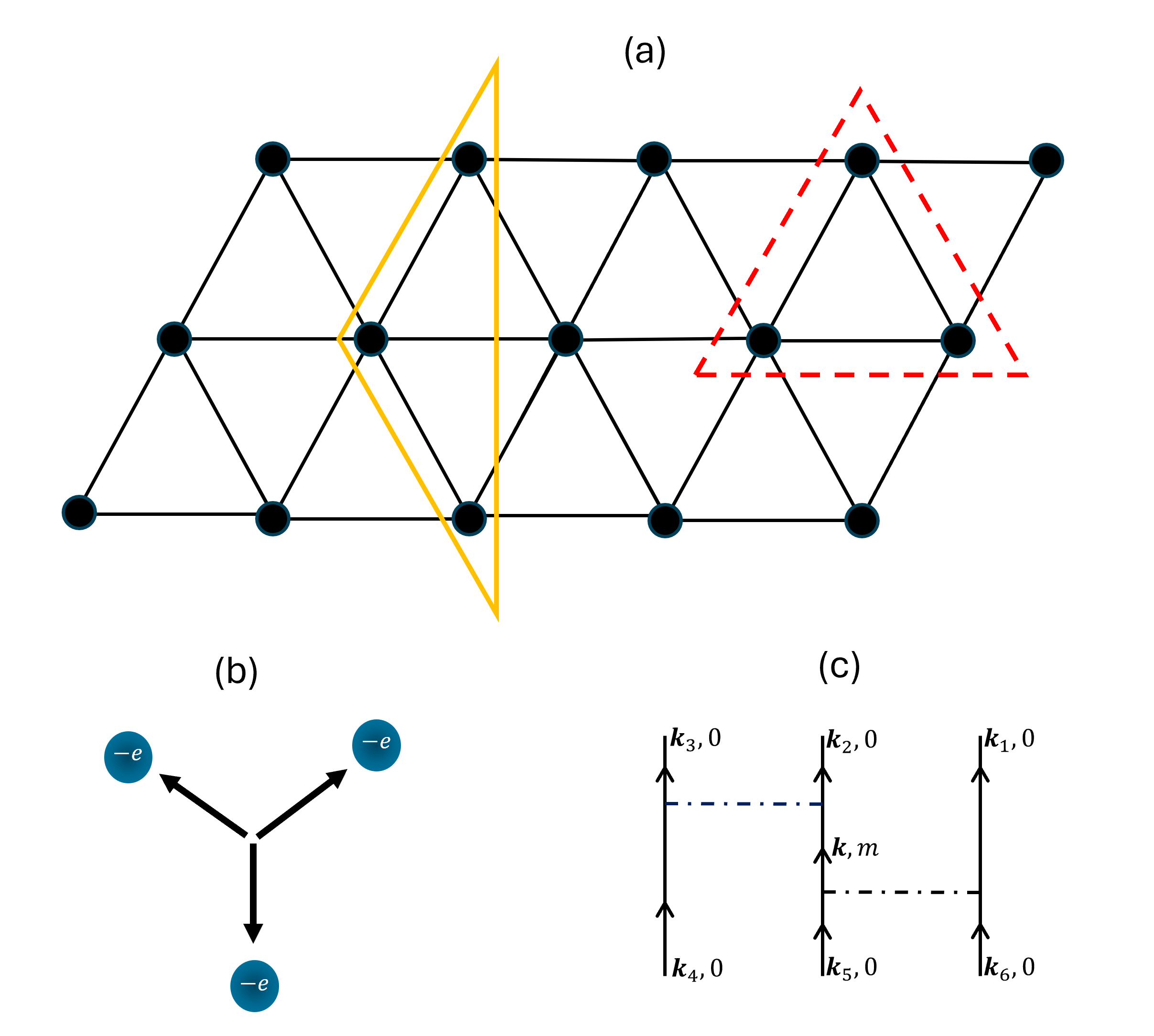}
    \caption{(a) Graphic representation of the nearest three electrons in a triangular lattice. There are two ways(red and orange triangular) that three electrons are close to each other. (b) When three electrons come close to each other, they experience repulsive force. (c) One of the channels of the scattering process for the three-body interaction is due to the band mixing of the Coulomb interaction. The index 0, and $m$ denote the Chern band and other bands, respectively.}
    \label{fig:3body}
\end{figure}
The Moore-Read state is the exact ground state of the three-body interaction in the QHS \cite{wojs2010global}. Our work generalizes the statement to the case of the Chern band system and explores its effects in a realistic setting.
Exact diagonalization is used to demonstrate the ground state of the Hamiltonian being in the same phase as the Moore-Read Pfaffian state. Energy spectrum, counting of excited states, and spectral flow are all consistent with those characteristic of Pfaffian state in Landau level systems and lattice models\cite{BernevigZoology,ShengNonabelian,BernevigTranslation}.
We investigate the competition between the composite Fermi liquid and the paired Pfaffian state by interpolating between the Coulomb interaction and the three-body interaction in a finite-size system.
The excited states in the energy spectrum will merge into the ground state manifold when increasing the Coulomb potential to a critical strength, which signals a phase transition in the thermodynamic limit. 
Furthermore, we make use of the perturbation theory to investigate the feasibility of this three-body interaction in a realistic system by considering the band mixing effect.

\begin{figure}[h]
    \centering
    \includegraphics[width=1.0\linewidth]{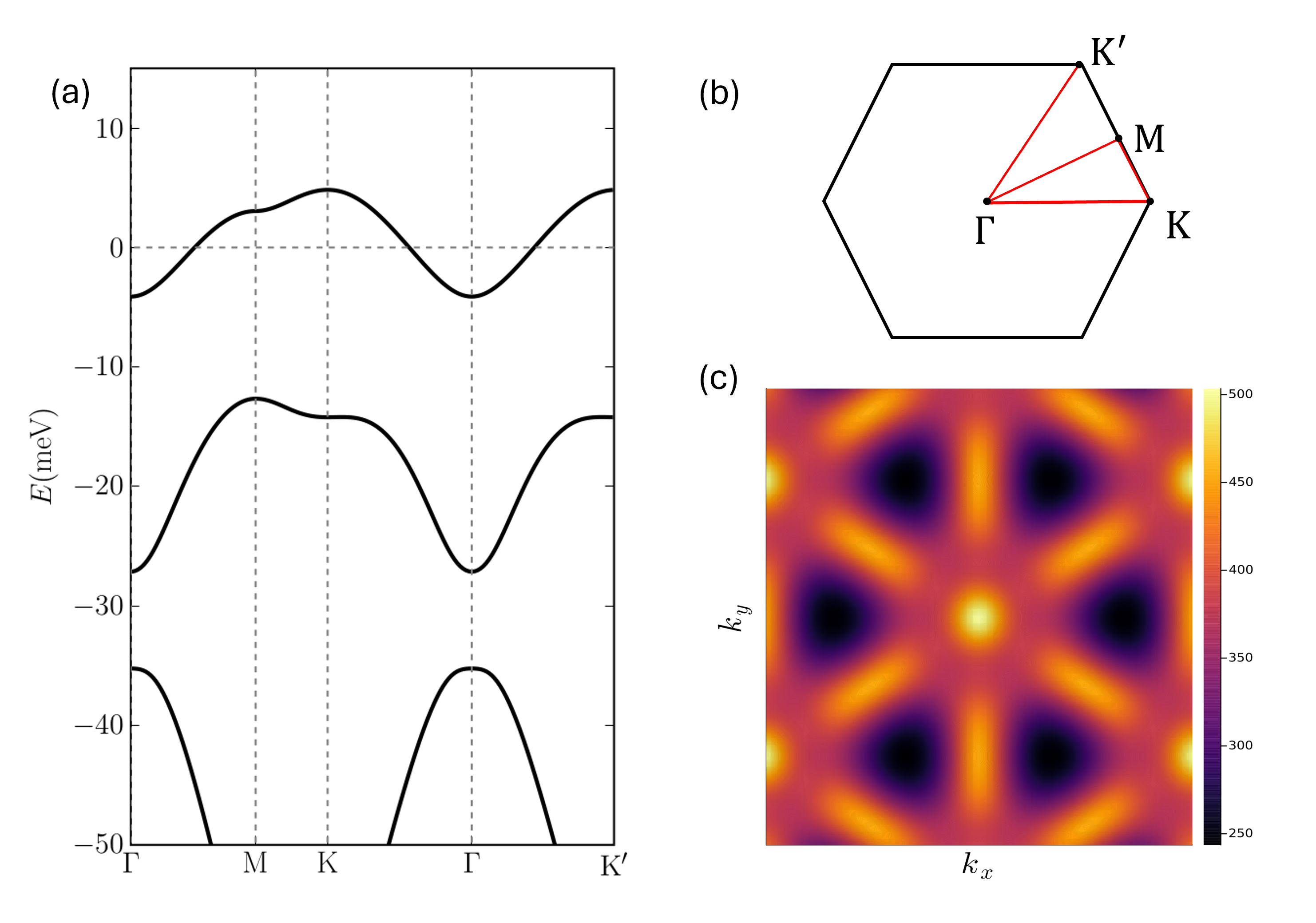}
    \caption{(a) The band structure of $\rm MoTe_2$ at twist angle $3.89^\circ$.  (b) The path along the high symmetry points in the first Brillouin zone. (c) The Berry curvature of the Chern band. }
    \label{fig:band}
\end{figure}

\textit{Formulation}.---
The toy Hamiltonian of the system incorporating the Coulomb interaction and the short-range three-body interaction is 
\begin{equation}  
    \hat{H} = \hat{H}_0 + \sum_{ij}  
    V^{\rm C}_{ij}:\hat \rho_{i}\hat \rho_{j}:
           + \sum_{ijk}V^{\rm 3b}_{ijk} :\hat \rho_{i} \hat \rho_{j} \hat \rho_{k}:
\end{equation}
where $\hat H_0$ is the single particle Hamiltonian of the continuum model\cite{wu2019topological} of the twisted $\rm MoTe_2$. $V^{\rm C}_{ij}$ and $V^{\rm 3b}_{ijk}$ denote the strength of the Coulomb interaction and the nearest three-body interaction respectively and $\hat \rho_i$ is the density operator. The normal ordering :: is employed to remove the self-interaction. 
We adopt the continuum model for a low energy description of the system at the single particle level. The Hamiltonian in one of the valleys($K$,$K^\prime$, which are locked with spins)\cite{wu2019topological}  is
\begin{equation}
    \hat{H}_0^{K \uparrow} = \left(\begin{array}{cc}
             \frac{\left(\boldsymbol{p} - \hbar \boldsymbol{K}_{t}\right)^2}{2m^{\star}}+\Delta_t(\boldsymbol{r}) & T(\boldsymbol{r}) \\
             T^{\dagger}(\boldsymbol{r})        & \frac{\left(\boldsymbol{p} - \hbar \boldsymbol{K}_{b}\right)^2}{2m^{\star}}+\Delta_b(\boldsymbol{r})
          \end{array}\right)
\end{equation}
where t/b denotes the top/bottom layer respectively. $m^*$ is the renormalized mass of the electron. The moir\'e potential
$\Delta_{t/b}=2V\sum_{j=1,3,5}\cos{(\boldsymbol{b}_j \cdot \boldsymbol{r} \pm \psi)}$ and interlayer hopping $T(\boldsymbol{r})=\omega (1+e^{i \boldsymbol{b}_2\cdot r}+e^{i \boldsymbol{b}_3\cdot r})$ where $V$ and $\psi$ determine the moir\'e potential and $\omega$ determines the strength of the interlayer hopping. $\boldsymbol{b}_j$ are the reciprocal vectors of the moir\'e lattice by rotating $\boldsymbol{b}_1$ by $(j-1)\pi/3$ counterclockwise. 
The Coulomb interaction between the valleys breaks their degeneracy and results in valley (spin) polarization\cite{devakul2021magic, li2021spontaneous, crepel2023fci}. 
Thus only one valley is considered here. Here we adopt the  first principle parameters ($V=20.8\rm{meV}$, $\psi=107.7^\circ$, $w=-23.8\rm{meV}$,  $m^{*}=0.62m_e$, $a_0=3.52\rm{\AA}$) from ref \onlinecite{XiaodiFractional}.
The twist angle we focus on is $3.89^\circ$, which is experimentally accessible.
The band structure and the berry curvature are shown in Fig. \ref{fig:band}(a,c). Numerical works have shown that the ideal band geometry is satisfied under this set of parameters, such that the flat Chern band can be effectively mapped to the lowest Landau level\cite{XiaodiFractional,wang2021exact,ledwith2023vortexability,ozawa2021relations}. The calculation using another set of parameters in Ref \cite{LiangFuFQAH}, supplemented with three-body interaction, also supports the Pfaffian states, as shown in Appendix \ref{appendix:otherparameter}.

The Coulomb interaction in Fourier space takes the form $V(\mathbf q)=\frac{e^2tanh(|q|d)}{ 2\epsilon_0 \epsilon_r|q|}$ with $\epsilon_{0(r)}$ the vacuum(relative) permittivity, which incorporates the gate-screening effect. We consider the short-range three-body repulsive interaction in single layer $\rm MoTe_2$. When three electrons are near each other, they experience repulsive forces shown in Fig.\ \ref{fig:3body}(b). 
The three-body repulsion in real space takes the form $ V^{\rm 3b}\sum_{i\neq j} \sum_{\boldsymbol{a},\boldsymbol{b}}\delta(\boldsymbol{r}_i+\boldsymbol{a},\boldsymbol{r}_j)\delta(\boldsymbol{r}_i+\boldsymbol{b},\boldsymbol{r}_k)$ where $\boldsymbol{r}_i,\boldsymbol{r}_j,\boldsymbol{r}_k$ are the positions of  three arbitrary electrons. $\boldsymbol a$ and $\boldsymbol b$ are vectors for the nearest neighbor bonds, and are summed over all pairs of different nearest neighbor bonds on the triangular lattice. There are two types of plaquettes on which three particles are close to each other as enclosed by red and yellow triangular in Fig.\ \ref{fig:3body}(a). Both of the plaquettes are incorporated in our calculation.
We next show that the ground state of the system is the Moore-Read state when the above three-body interaction dominates. The mechanism behind this phenomenon is rooted in the fact that the Moore-Read Pfaffian state is a paired state. Intuitively, the three-body short-range repulsion penalizes states in which three particles come closer together, while two particles coming together does not result in an energy cost.
By multiple trials, we find the low-energy theory quite robust and insensitive to the details of the three-body interaction.
The nearest neighbor three-body interaction we choose is just the simplest three-body interaction consistent with the symmetry of the system.
Therefore, our result from the point-contact three-body interaction is general and can be generalized to other short-range three-body interaction forms.

\textit{Energy spectrum, spectral flow and excited state counting}.---
To demonstrate the theoretical existence of the Moore-Read state in moir\'e systems, we set the strength of the three-body interaction to be comparable with the bare Coulomb interaction, and the dielectric constant $\epsilon_r=20$, which results in a relatively small Coulomb interaction. The low energy properties of the system are governed mainly by the three-body interaction. We project the three-body interaction onto the Chern band and perform exact diagonalization to obtain the groundstates and energy spectrum of the above Hamiltonian. We evaluate it on the torus with three different system sizes  $4\times4$, $4\times5$ and $4\times6$, respectively. 
\begin{figure}[h]
    \centering
    \includegraphics[width=1.0\linewidth]{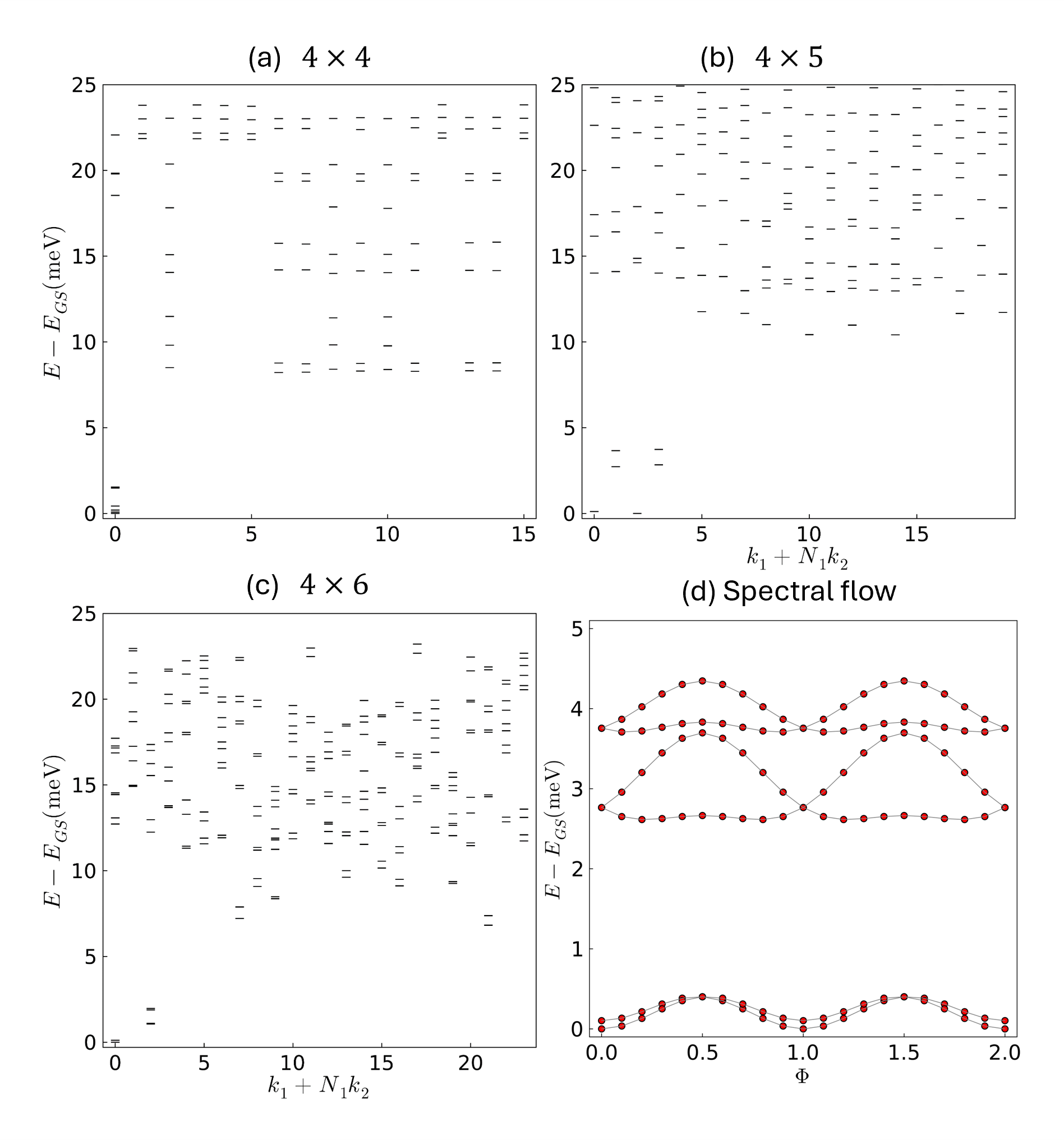}
    \caption{(a-c) The energy spectrum of half filling of twisted $\rm MoTe_2$ for three different size $4\times 4$, $4\times 5$, $4\times 6$, respectively. The dielectric constant $\epsilon_r=20$. There is a clear energy gap between the six-fold ground state and the excited states. (d) The spectral flow of the ground state wave function by inserting magnetic flux along $k_y$ direction.}
    \label{fig:spectrum}
\end{figure}
The sixfold degeneracy of the Moore-Read Pfaffian state on the torus is a well-known characteristic of this state from analysis of the topological field theory\cite{RMPnonabelian,oshikawa2007topological}. The presence of a gap between the ground state and excited state, along with the observed six-fold quasi-degeneracy of the ground state at different system sizes in Fig.\ \ref{fig:spectrum}(a-c), strongly suggests that the ground state is the Moore-Read Pfaffian state. Due to the finite size effect, there is a degeneracy lift\footnote{We note that the dielectric constant in TMDs falls in a range of $10\sim 20$, while we take $20$ here. The specific value one uses should not qualitatively alter the results, so long as three-body interaction dominates Coulomb interaction strength, a Pfaffian state regime will exist.}. Since the exact diagonalization only involves the localized interaction it is natural to expect that the Moore-Read state exists in the thermodynamic limit of the model. We further check the claim by inserting the flux through one cycle of the toturs. We find that the ground states flow into each other and return to itself after $2$ flux quanta without merging with higher energy states [cf. Fig.\ \ref{fig:spectrum}(d)].  

\begin{figure}[h]
    \centering
    \includegraphics[width=1\linewidth]{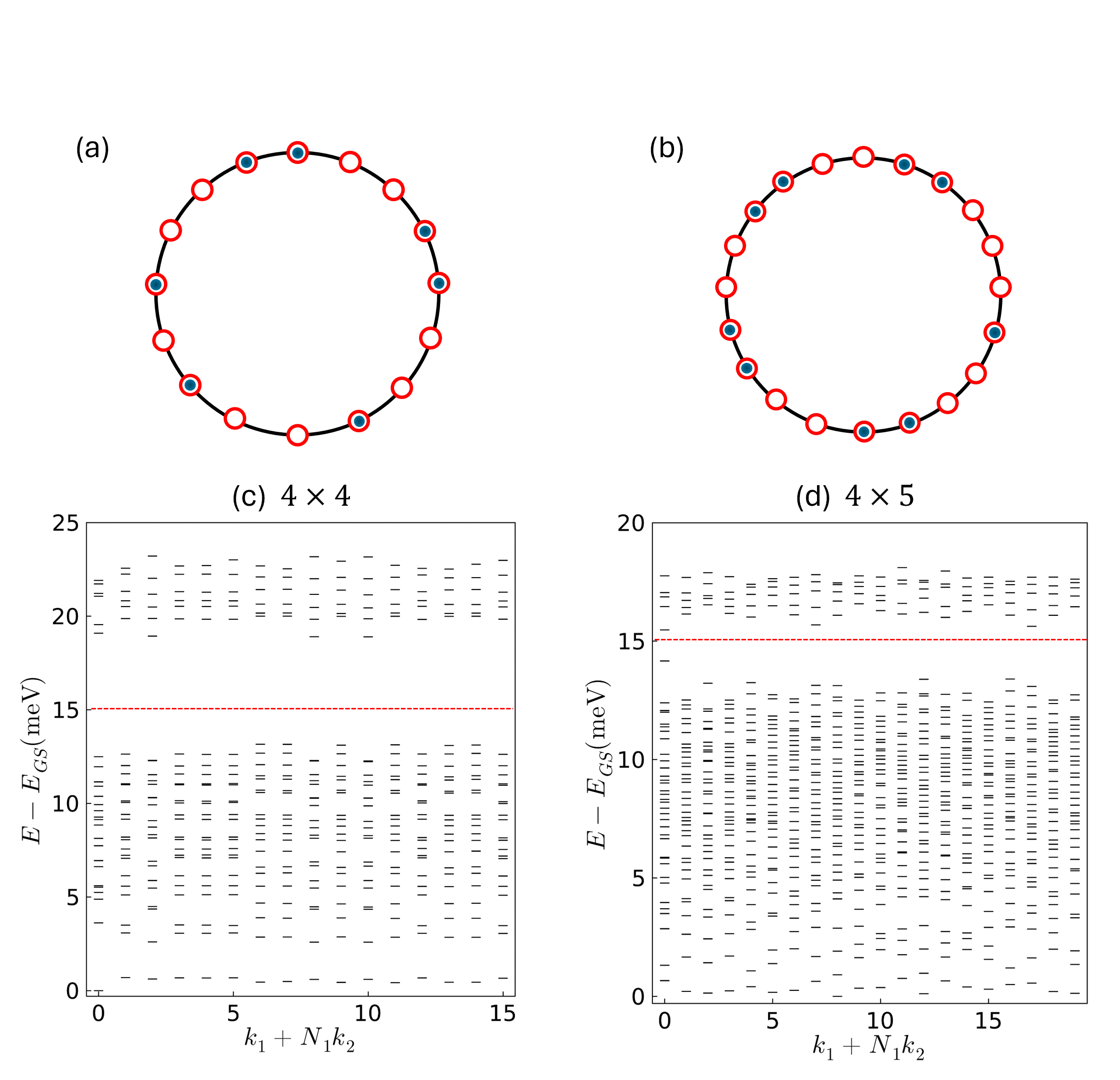}
    \caption{(a-b) The possible configurations of low energy states with one hole that satisfy the generalized Pauli principle corresponding to two system sizes 4x4 and 4x5. (c-d) The spectrum of the system that deviates from half-filling. The system size for (c) is 4x4 with 7 electrons and for (d) is $4\times 5$ with 9 electrons. States below the red dotted line are the low energy excitation.}
    \label{fig:excitation}
\end{figure}

To further confirm the non-Abelian properties, we count the states within the low-energy manifold by removing an electron and creating quasiholes above the 6-fold degenerate ground state \cite{ShengNonabelian}. The low-energy states must adhere to the generalized Pauli exclusion principle, which stipulates that no more than two electrons can occupy four consecutive orbitals((2,4) admissible)\cite{BernevigFCI}. According to combinatorics it is found that the number of low-energy states(below the red dotted line) should be 320 for 16 orbitals with 7 electrons and 700 for 20 orbitals with 9 electrons, which is consistent with our calculation  [cf. Fig.\ \ref{fig:excitation}]. The number of low energy excitations in each momentum sector can also be obtained by considering the center of mass and relative translational symmetry\cite{BernevigTranslation}.

\textit{Coulomb interaction and Competition with composite Fermi liquid}.---In real materials, the effect of Coulomb interaction cannot be neglected, which would compete with the three-body interaction. Previous studies have shown that the Coulomb interaction in the moir\'e system stabilizes the anomalous composite Fermi liquid, which is gapless\cite{LiangFuCFL,AshvinCFL}. To reveal the competition between the Moore-Read state and the composite Fermi liquid and how likely it is to realize the Moore-Read state, we interpolate between the three-body interaction and the Coulomb interaction.
\begin{equation}
    \hat{H}_{\rm{int}}=\hat{H}_0+\alpha \hat{V}^{\rm{C}}+(1-\alpha)\hat{V}^{\rm{3b}}
\end{equation}
where the $\hat{V}^C$ is the Coulomb interaction term and the $\hat{V}^{\rm{3b}}$ is three-body interaction term. The strengths $V^{\rm 3b}$ is set to be $\frac{e^2}{\epsilon_0 a_0}$. For numerical efficiency, we perform the calculation on the $4\times 4$ torus. It is expected that as $\alpha$ approaches $1$, the gap will close and the ground state will be the anomalous composite Fermi liquid state. 
\begin{figure}[h]
    \centering
    \includegraphics[width=1.\linewidth]{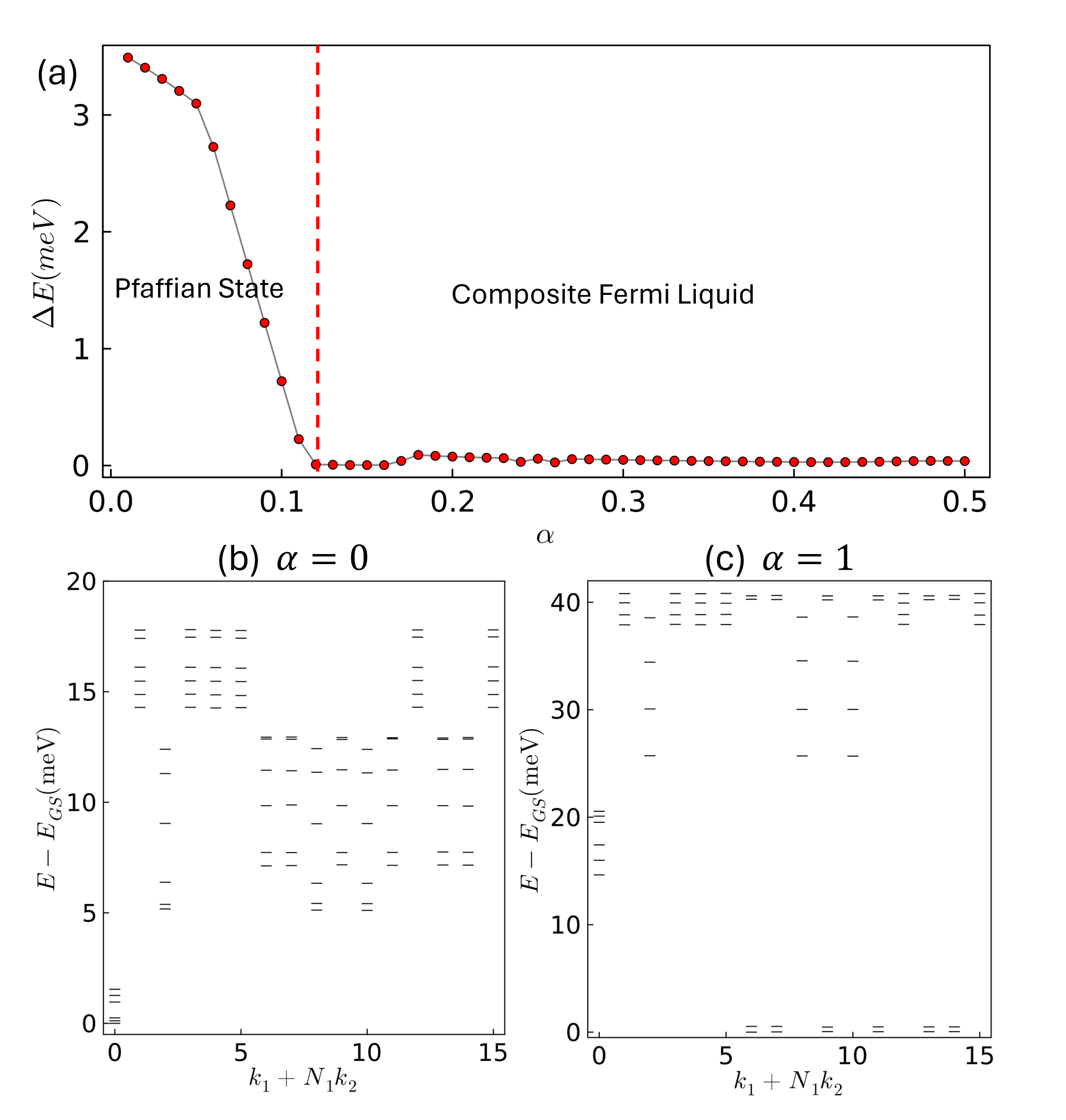}
    \caption{(a) Interpolation between the Coulomb interaction and the three-body interaction which shows the gap of the lowest excited state as a function of the interpolation parameter $\alpha$. (b-c) The spectrum at two extreme cases corresponding to $\alpha = 0$ and $\alpha = 1$.}
    \label{fig:interpolation}
\end{figure}
When the interpolation parameter $\alpha$ is set to $0$, the system, dominated by the three-body interaction, stays in an incompressible state and thus the gap remains. As $\alpha$ increases to around 0.12, the gap closes, indicating a phase transition\ [cf. Fig.\ \ref{fig:interpolation}(a)]. The system moves from the Moore-Read state to the composite Fermi liquid. The spectrum of two extreme cases $\alpha=0$ and $\alpha=1$ are shown in Fig.\ \ref{fig:interpolation}(b) and Fig.\ \ref{fig:interpolation}(c), respectively. Both results are consistent with our calculations and other studies. From the above analysis, we conclude that a relatively large three-body repulsive interaction is necessary for the Pfaffian state to be the ground state of the system. The fact that the gap continuously closes from the Pfaffian state to the CFL is consistent with a continuous phase transition proposed previously\cite{metlitski2015cooper,WANG2014pairing}.

\textit{Three-body interaction from band mixing}.---It is natural to ask whether it is possible to realize three-body interaction in a realistic platform that leads to the Moore-Read state. We show that the three-body interaction can arise due to the band mixing and determine the contribution of the band mixing to the three-body interaction using perturbation theory. Dividing the Coulomb interaction into two parts, the Hamiltonian takes the form
\begin{equation}
    \hat{H} = \hat{H}_0 + \hat P_0 \hat{V}^C \hat P_0 + \hat P_0 \hat{V}^C \hat P_1 + h.c.
\end{equation}
where $P_0$ and $P_1$ are the projection operator that projects the state to the Chern band and the other bands, respectively, and $\hat{H}_0$ is the single particle Hamiltonian. By the Schrieffer-Wolff transformation\cite{coleman2015introduction}, the interactions that induce inter-band scattering are renormalized into the interaction within the lowest Chern band. The renormalized interaction reads:
\begin{equation}
- \hat P_0 \hat{V}^{\rm{C}} \hat P_1 \frac{1}{\hat H_0 -E_0} \hat P_1 \hat{V}^{\rm{C}} \hat P_0 
\label{eq:sw}
\end{equation}
When $\hat P_0 \hat{V}^{\rm{C}} \hat P_1$ takes one of the electrons in the Chern band to other bands and leaves the other electron in the original band, the three-body interaction emerges\cite{MacDonaldmixing,Rezayimixing,Rezayimixing2},  which can be graphically represented in Fig.\ref{fig:3body}(c).  
From eq \eqref{eq:sw} we will see the effect of the band mixing is determined by: (1) The form factor $F\left(\boldsymbol{k}_1, \boldsymbol{k}_2,m,n, \boldsymbol{g}\right)=\sum_{\boldsymbol{g}^{\prime} l} \psi_{m\boldsymbol{k}_1\left(\boldsymbol{g}^{\prime}+\boldsymbol{g}\right) l}^* \psi_{n \boldsymbol{k}_2 \boldsymbol{g}^{\prime} l}$ where $\psi_{m,\boldsymbol{k}}$ denotes the Bloch wave function of momentum $\boldsymbol{k}$ in band $m$ with $l$ and $\boldsymbol{g,g'}$ the layer degrees of freedom and moir\'e reciprocal vectors, respectively. (2) The dielectric constant $\epsilon_r$ and (3) the band gap $\Delta E$. Through numerical calculations, we have determined that the magnitude of form factor $F(\boldsymbol{k}_1,\boldsymbol{k}_2,0,0,\boldsymbol{g})$ associated with electron scattering within the lowest Chern band and the form factor $F(\boldsymbol{k}_1,\boldsymbol{k}_2,0,m,\boldsymbol{g})$ for scattering out of the lowest Chern band are comparable, and approximate to $10^{-2}$.

For the short range three-body interaction to arise from band mixing effects, we propose one condition that leads to the Moore-Read state,
\begin{equation}
\begin{aligned}
    \frac{1}{N_{\boldsymbol{k}}}\sum_{m\neq 0,\boldsymbol{k}_s} F\left(\boldsymbol{k}_2, \boldsymbol{k}_s, 0, m, \boldsymbol{g}_1\right)&  F\left(\boldsymbol{k}_s, \boldsymbol{k}_5, m, 0, \boldsymbol{g}_2\right)\frac{1}{\Delta E_m} \\ &= C\ F\left(\boldsymbol{k}_2, \boldsymbol{k}_5, 0,0, \boldsymbol{g}_1+\boldsymbol{g}_2\right)
\end{aligned}
\label{eq:condition}
\end{equation}
where $N_{\boldsymbol{k}}$ is the number of $\boldsymbol{k}$ points proportional to the system area and $C$ is a constant with inverse energy dimension.
Note that this is not necessary condition, as other interaction forms may well favor Pfaffian states, which remain to be explored. The above formula poses a constraint of the form factor and simplifies the three-body interaction from perturbation as
\begin{small}
\begin{align}
V^{\rm{per}}_{\substack{\boldsymbol{k}_1, \boldsymbol{k}_2, \boldsymbol{k}_3\\
\boldsymbol{k}_4, \boldsymbol{k}_5, \boldsymbol{k}_6}} = C \sum_{\boldsymbol{g}_1,\boldsymbol{g}_2,\boldsymbol{g}_3}\delta_{\rm{ksum}}V(\boldsymbol{k}_1-\boldsymbol{k}_4+\boldsymbol{g}_1)V(\boldsymbol{k}_3-\boldsymbol{k}_6+\boldsymbol{g}_3) \times  \nonumber\\ 
F(\boldsymbol{k}_1,\boldsymbol{k}_4,0,0,\boldsymbol{g}_1)F(\boldsymbol{k}_2,\boldsymbol{k}_5,0,0,\boldsymbol{g}_2)F(\boldsymbol{k}_3,\boldsymbol{k}_6,0,0,\boldsymbol{g}_3)
\label{eq:3bper}
\end{align}
\end{small}
where $\delta_{\rm{ksum}} =\delta_{\boldsymbol{k}_1+\boldsymbol{k}_2+\boldsymbol{k}_3,\boldsymbol{k}_4+\boldsymbol{k}_5+\boldsymbol{k}_6+\boldsymbol{g}_1+\boldsymbol{g}_2+\boldsymbol{g}_3}$, and the left hand side is the matrix element of the three-body interaction $V^{\rm 3b}$. The detailed derivation is provided in the Appendix \ref{appendix:3b}. 

As a crude estimate of the strength of three-body interaction from band mixing, we note that the projected Coulomb interaction is of order $\frac{e^2}{\epsilon_0 \epsilon_r a_M} \sim 10^2 \rm{meV}$, while the three-body interaction from eq \eqref{eq:sw} is $(\frac{e^2}{\epsilon_0 \epsilon_r a_M})^2/(\Delta E)\sim 10^3 \rm{meV}$ weighted by  the form factor $F$ from the inter-band scattering, of order $0.05$. So solely from band mixing mechanism, we have in the lowest Chern band $V^{\rm C}/V^{\rm 3b}\sim 2$, which is 
not in but not far from the Pfaffian state regime in Fig \ref{fig:interpolation}. However, we note it is very crude and does not take into account the condition of eq \eqref{eq:condition} or the detailed form of $F$ funcion carefully.

\textit{Summary and discussion}.--- In conclusion, we have numerically demonstrated the existence of the Moore-Read state at half-filling upon introducing a three-body interaction to a moir\'e Chern band model, specifically twisted bilayer $\rm MoTe_2$. As shown in the phase diagram Fig.\ref{fig:interpolation}(a), when the three-body interaction becomes dominant, it leads to the non-Abelian state. Additionally, we have shown that the three-body interaction can emerge as a low-energy effect from band mixing by perturbation theory. The engineering of such interactions in physical systems remains an important issue to resolve in the future. Looking forward, it would be interesting to explore the connection to the putative fractional quantum spin hall states observed in moir\'e TMDs \cite{kang2024observation}, and ask whether the interplay of valley and inter-band scattering could stabilize an non-Abelian state there.
It would be interesting to pursue realizations of other non-Abelian states, such as the Read-Rezayi state, in moir\'e systems. The search for non-Abelian fractionalization with more feasible physical conditions in tunable platforms, such as wide quantum well, (twisted) graphene multilayers and strained graphene, etc\cite{gao2023untwisting,ledwith2022family,ledwith2020fractional, dong2023many} are also worth exploring.

\textit{Acknowledgements:} We thank Aidan Reddy for discussions and generously sharing part of the \href{http://github.com/AidanReddy/FermionED/tree/main}{codes on exact diagonalization}. We thank Ahmed Abouelkomsan, Junkai Dong, Patrick Ledwith, Todadri Senthil, and Ya-Hui Zhang for helpful discussions.
XYS was supported by the Gordon and Betty Moore Foundation EPiQS Initiative through Grant No.~GBMF8684 at the Massachusetts Institute of Technology.

\nocite{1}

\input{Main.bbl}
\onecolumngrid 

\appendix
\section{Energy spectrum using other first principle parameters}
\label{appendix:otherparameter}
\begin{figure}[h]
    \centering
    \includegraphics[width=1\linewidth]{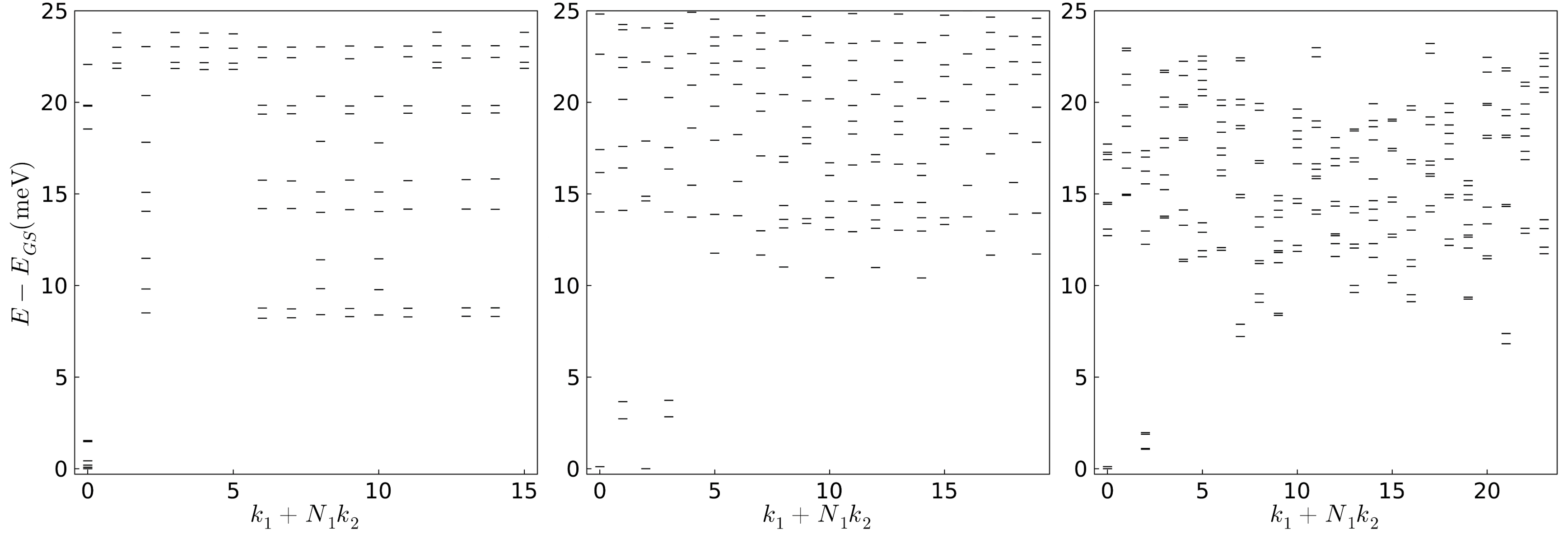}
    \caption{the energy spectrum of half filling of twisted
             $\rm{MoTe}_2$ for three different size $4\times 4, 4\times 5 , 4\times 6$. }
    \label{fig:spectrumLiangFu}
\end{figure}

There are different sets of first principle calculation parameters for the twisted bilayer TMDC. In this section, we use the parameters $V=11.2\rm{meV}$, $\phi=-91^\circ$, $w=-11.3\rm{meV}$,  $m^{*}=0.62m_e$, $a_0=3.52\rm{\AA}$)\cite{LiangFuFQAH} at twist angle 
 $\theta = 1.9^\circ$. However, the ratio between the three-body and the Coulomb interaction should be larger than the ratio in the main text for the ground state to be the Moore-Read state. We choose $\epsilon_r=20$ and the three-body interaction strength $V^{\rm 3b}=\frac{15e^2}{\epsilon_0 a_0}$. The energy spectrum is shown in the Fig.\ \ref{fig:spectrumLiangFu}

The energy gap and the ground state degeneracy indicate that the ground state is the Moore-Read state. All other results in the main body can be reproduced using this set of parameters.

\section{three-body Interaction Derived by Perturbation Theory}
\label{appendix:3b}

The following effective Hamiltonian can be constructed by considering the effective theory:
\begin{equation}
    \hat H_{\mathrm{eff}}=\hat P_0\hat H_0 \hat P_0+ \hat P_0 \hat V \hat P_0 -\hat P_0 \hat V \hat P_{1} \frac{1}{\hat H_0-E_0} \hat P_{1} \hat V \hat P_0
\end{equation}
where $\hat P$ is the project operator that projects the system to the Chern band and $\hat H$ is the single particle Hamiltonian. The first order term $\hat H_1 = \hat P_0 \hat V \hat P_0$ of the perturbation is simply the projected Coulomb interaction. Now we calculate the second order term $\hat H_2 = \hat P_0 \hat V \hat P_{1} \frac{1}{\hat H_0-E_0} \hat P_{1} \hat V \hat P_0$
In the following we use $m$ to denote the index of the band and 0 refers to the Chern band projected onto. Write the second-order perturbation term in terms of the creations and annihilation operators, which diagonalize the Chern band.
\begin{equation}
\begin{aligned}
\hat H_2 = & \sum_{\substack{\boldsymbol{k}_1, \boldsymbol{k}_2, \boldsymbol{k}_3, \boldsymbol{k}_4\\\boldsymbol{k}_1^\prime, \boldsymbol{k}_2^\prime, \boldsymbol{k}_3^\prime, \boldsymbol{k}_4^\prime\\ m_1,m_2,m_3,m_4}} \left(V_{\substack{\boldsymbol{k}_1, \boldsymbol{k}_2, \boldsymbol{k}_3, \boldsymbol{k}_4\\0,0, m_3, m_4}} a_{0, \boldsymbol{k}_1}^{\dagger} a_{0, \boldsymbol{k}_2}^{\dagger} a_{m_3, \boldsymbol{k}_3} a_{m_4, \boldsymbol{k}_4}\right) \hat P_1 \frac{1}{\hat H_0-E_0} \hat P_1\left(V_{\substack{\boldsymbol{k}_1^{\prime}, \boldsymbol{k}_2^\prime, \boldsymbol{k}_3^\prime, \boldsymbol{k}_4^{\prime}\\ m_1, m_2,0,0}} a_{m_1, \boldsymbol{k}_1^{\prime}}^{\dagger} a_{m_2, \boldsymbol{k}_2^{\prime}}^{\dagger} a_{0, \boldsymbol{k}_3^{\prime}} a_{0, \boldsymbol{k}_4^{\prime}}\right)  \\
\end{aligned}
\end{equation}
where $m$ is the index of the bands and cannot equal 0.  The operator $\hat P_1 \hat V \hat P_0$ will scatter the electrons from the Chern band to higher bands. When only one electron in the Chern band is scattered into the Chern band, the three-body interaction emerges. When two electrons are scattered, there is an additional two-body interaction.  We assume the two-body interaction is still governed by the Coulomb interaction.

Choosing the relative dielectric constant $\epsilon_r=10$, the energy scale of the Coulomb interaction is $\frac{e^2}{\epsilon_0 \epsilon_r a_M} \sim 10^2 \rm{meV}$. The strength of the three-body interaction derived from the perturbation is approximately $(\frac{e^2}{\epsilon_0 \epsilon_r a_M})^2/(\Delta E)\sim 10^3 \rm{meV}$. 
The ratio between the strength of the three-body interaction and the Coulomb interaction ($\sim 10$) makes it possible to realize the Moore-Read state in the system as shown in our result in Fig.\ \ref{fig:interpolation}.
The energy scale of the band mixing effect also depends on the form factor $F\left(\boldsymbol{k}_1, \boldsymbol{k}_2, 0, m, \boldsymbol{g}\right)$, which is small ($\sim 0.05$) and suppresses the numerical value of the term.
Hence perturbation theory holds here.

In our case, we only focus on the three-body interaction. 
In this case one of the creation operators $\left\{ a_{m_1, \boldsymbol{k}_1^{\prime}}^{\dagger}, a_{m_2, \boldsymbol{k}_2^{\prime}}^{\dagger} \right\}$ scatters electrons into the higher bands and one of the annihilation operators $\left\{ a_{m_3, \boldsymbol{k}_3}, a_{m_4, \boldsymbol{k}_4} \right\}$ bring it back. There are four cases in total.
\begin{enumerate}
    \item $\boldsymbol{k}_1^{\prime}=\boldsymbol{k}_3$ , $m_1 = m_3$ and $m_2 = m_4 =0$\\
\begin{equation}
\begin{aligned}
\hat H_2 = & \sum_{\substack{\boldsymbol{k}_1, \boldsymbol{k}_2, \boldsymbol{k}_4\\ \boldsymbol{k}_2^\prime, \boldsymbol{k}_3^\prime, \boldsymbol{k}_4^\prime}}\sum_{m,\boldsymbol{k}} \left(V_{\substack{\boldsymbol{k}_1, \boldsymbol{k}_2, \boldsymbol{k}, \boldsymbol{k}_4\\0,0, m, 0}} a_{0, \boldsymbol{k}_1}^{\dagger} a_{0, \boldsymbol{k}_2}^{\dagger} a_{m, \boldsymbol{k}} a_{0, \boldsymbol{k}_4}\right) \hat P_1 \frac{1}{\hat H_0-E_0} \hat P_1\left(V_{\substack{\boldsymbol{k}_1^{\prime}, \boldsymbol{k}_2^\prime, \boldsymbol{k}_3^\prime, \boldsymbol{k}_4^{\prime}\\ m, 0,0,0}} a_{m, \boldsymbol{k}}^{\dagger} a_{0, \boldsymbol{k}_2^{\prime}}^{\dagger} a_{0, \boldsymbol{k}_3^{\prime}} a_{0, \boldsymbol{k}_4^{\prime}}\right)  \\
=& \sum_{\substack{\boldsymbol{k}_1, \boldsymbol{k}_2, \boldsymbol{k}_4\\ \boldsymbol{k}_2^\prime, \boldsymbol{k}_3^\prime, \boldsymbol{k}_4^\prime}}\sum_{m,\boldsymbol{k}}V_{\substack{\boldsymbol{k}_1, \boldsymbol{k}_2, \boldsymbol{k}, \boldsymbol{k}_4\\0,0, m, 0}}V_{\substack{\boldsymbol{k}, \boldsymbol{k}_2^\prime, \boldsymbol{k}_3^\prime, \boldsymbol{k}_4^{\prime}\\ m, 0,0,0}} \left( a_{0, \boldsymbol{k}_1}^{\dagger} a_{0, \boldsymbol{k}_2}^{\dagger} a_{m, \boldsymbol{k}} a_{0, \boldsymbol{k}_4}\right) \hat P_1 \frac{1}{\hat H_0-E_0} \hat P_1\left( a_{m, \boldsymbol{k}}^{\dagger} a_{0, \boldsymbol{k}_2^{\prime}}^{\dagger} a_{0, \boldsymbol{k}_3^{\prime}} a_{0, \boldsymbol{k}_4^{\prime}}\right)\\
=& \sum_{\substack{\boldsymbol{k}_1, \boldsymbol{k}_2, \boldsymbol{k}_4\\ \boldsymbol{k}_2^\prime, \boldsymbol{k}_3^\prime, \boldsymbol{k}_4^\prime}}\left(\sum_{m,\boldsymbol{k}} \frac{-1}{\Delta E_m} V_{\substack{\boldsymbol{k}_1, \boldsymbol{k}_2, \boldsymbol{k}, \boldsymbol{k}_4\\0,0, m, 0}}V_{\substack{\boldsymbol{k}, \boldsymbol{k}_2^\prime, \boldsymbol{k}_3^\prime, \boldsymbol{k}_4^{\prime}\\ m, 0,0,0}}\right) \left( a_{0, \boldsymbol{k}_1}^{\dagger} a_{0, \boldsymbol{k}_2}^{\dagger}a_{0, \boldsymbol{k}_4} a_{0, \boldsymbol{k}_2^{\prime}}^{\dagger} a_{0, \boldsymbol{k}_3^{\prime}} a_{0, \boldsymbol{k}_4^{\prime}}\right)\\
\end{aligned}
\end{equation}
Where $V_{{\substack{\boldsymbol{k}_1,\boldsymbol{k}_2,\boldsymbol{k}_3,\boldsymbol{k}_4\\m_1,m_2,m_3,m_4}}}= \frac{1}{A}\sum_{\boldsymbol{q}} \delta_{\boldsymbol{k}_3-\boldsymbol{q},\boldsymbol{k}_2+\boldsymbol{g}_1} \delta_{\boldsymbol{k}_4+\boldsymbol{q},\boldsymbol{k}_1+\boldsymbol{g}_2} V(\boldsymbol{q})F\left(\boldsymbol{k}_2, \boldsymbol{k}_3, m_2, m_3, \boldsymbol{g}_1\right) F\left(\boldsymbol{k}_1, \boldsymbol{k}_4, m_1, m_4, \boldsymbol{g}_2\right)$. In the second equality we approximate $\frac{1}{\hat H_0 -E_0}$ as $\frac{1}{\Delta E_m}$ . We normal ordered the above formula to obtain the three-body interaction:
\begin{equation}
 \sum_{\substack{\boldsymbol{k}_1, \boldsymbol{k}_2, \boldsymbol{k}_4\\ \boldsymbol{k}_2^\prime, \boldsymbol{k}_3^\prime, \boldsymbol{k}_4^\prime}}\left(\sum_{m,\boldsymbol{k}} \frac{1}{\Delta E_m} V_{\substack{\boldsymbol{k}_1, \boldsymbol{k}_2, \boldsymbol{k}, \boldsymbol{k}_4\\0,0, m, 0}}V_{\substack{\boldsymbol{k}, \boldsymbol{k}_2^\prime, \boldsymbol{k}_3^\prime, \boldsymbol{k}_4^{\prime}\\ m, 0,0,0}}\right) \left( a_{0, \boldsymbol{k}_1}^{\dagger} a_{0, \boldsymbol{k}_2}^{\dagger} a_{0, \boldsymbol{k}_2^{\prime}}^{\dagger} a_{0, \boldsymbol{k}_4}  a_{0, \boldsymbol{k}_3^{\prime}} a_{0, \boldsymbol{k}_4^{\prime}}\right)
\end{equation}
Using the same trick we obtain the remaining three cases.
    \item $\boldsymbol{k}_2^{\prime}=\boldsymbol{k}_3$ , $m_2 = m_3$ and $m_1 = m_4 =0$\\ 
\begin{equation}
 \sum_{\substack{\boldsymbol{k}_1, \boldsymbol{k}_2, \boldsymbol{k}_4\\ \boldsymbol{k}_1^\prime, \boldsymbol{k}_3^\prime, \boldsymbol{k}_4^\prime}}\left(\sum_{m,\boldsymbol{k}} \frac{-1}{\Delta E_m} V_{\substack{\boldsymbol{k}_1, \boldsymbol{k}_2, \boldsymbol{k}, \boldsymbol{k}_4\\0,0, m, 0}}V_{\substack{\boldsymbol{k}_1^\prime, \boldsymbol{k}, \boldsymbol{k}_3^\prime, \boldsymbol{k}_4^{\prime}\\ 0, m,0,0}}\right) \left( a_{0, \boldsymbol{k}_1}^{\dagger} a_{0, \boldsymbol{k}_2}^{\dagger} a_{0, \boldsymbol{k}_1^{\prime}}^{\dagger} a_{0, \boldsymbol{k}_4}  a_{0, \boldsymbol{k}_3^{\prime}} a_{0, \boldsymbol{k}_4^{\prime}}\right)
\end{equation}
    \item $\boldsymbol{k}_1^{\prime}=\boldsymbol{k}_4$ , $m_1 = m_4$ and $m_2 = m_3 =0$\\ 
\begin{equation}
 \sum_{\substack{\boldsymbol{k}_1, \boldsymbol{k}_2, \boldsymbol{k}_3\\ \boldsymbol{k}_2^\prime, \boldsymbol{k}_3^\prime, \boldsymbol{k}_4^\prime}}\left(\sum_{m,\boldsymbol{k}} \frac{-1}{\Delta E_m} V_{\substack{0,0,\boldsymbol{k}_1, \boldsymbol{k}_2, \boldsymbol{k}_3, \boldsymbol{k}\\ 0, m}}V_{\substack{\boldsymbol{k}, \boldsymbol{k}_2^\prime, \boldsymbol{k}_3^\prime, \boldsymbol{k}_4^{\prime}\\ m, 0,0,0}}\right) \left( a_{0, \boldsymbol{k}_1}^{\dagger} a_{0, \boldsymbol{k}_2}^{\dagger} a_{0, \boldsymbol{k}_2^{\prime}}^{\dagger} a_{0, \boldsymbol{k}_3}  a_{0, \boldsymbol{k}_3^{\prime}} a_{0, \boldsymbol{k}_4^{\prime}}\right)
\end{equation}
    \item $\boldsymbol{k}_2^{\prime}=\boldsymbol{k}_4$ , $m_2 = m_4$ and $m_1 = m_3 =0$\\ 
\begin{equation}
 \sum_{\substack{\boldsymbol{k}_1, \boldsymbol{k}_2, \boldsymbol{k}_3\\ \boldsymbol{k}_1^\prime, \boldsymbol{k}_3^\prime, \boldsymbol{k}_4^\prime}}\left(\sum_{m,\boldsymbol{k}} \frac{1}{\Delta E_m} V_{\substack{\boldsymbol{k}_1, \boldsymbol{k}_2, \boldsymbol{k}_3, \boldsymbol{k}\\0,0, 0, m}}V_{\substack{\boldsymbol{k}_1^\prime, \boldsymbol{k}, \boldsymbol{k}_3^\prime, \boldsymbol{k}_4^{\prime}\\ 0, m,0,0}}\right) \left( a_{0, \boldsymbol{k}_1}^{\dagger} a_{0, \boldsymbol{k}_2}^{\dagger} a_{0, \boldsymbol{k}_1^{\prime}}^{\dagger} a_{0, \boldsymbol{k}_3}  a_{0, \boldsymbol{k}_3^{\prime}} a_{0, \boldsymbol{k}_4^{\prime}}\right)
\end{equation}    
\end{enumerate}
Thus the strength of the three-body interaction can be expressed as:
\begin{equation}
    V^{\rm{per}}_{\substack{\boldsymbol{k}_1, \boldsymbol{k}_2, \boldsymbol{k}_3, \\ \boldsymbol{k}_4, \boldsymbol{k}_5, \boldsymbol{k}_6}}= \sum_{m,\boldsymbol{k}}\frac{1}{\Delta E_m}\left( V_{\substack{\boldsymbol{k}_1, \boldsymbol{k}_2, \boldsymbol{k}, \boldsymbol{k}_4\\ 0,0,m,0}}V_{\substack{\boldsymbol{k}, \boldsymbol{k}_3, \boldsymbol{k}_5, \boldsymbol{k}_6\\ m, 0,0,0}}-V_{\substack{\boldsymbol{k}_1, \boldsymbol{k}_2, \boldsymbol{k}, \boldsymbol{k}_4\\ 0,0,m,0}}V_{\substack{\boldsymbol{k}_3, \boldsymbol{k}, \boldsymbol{k}_5, \boldsymbol{k}_6\\ 0, m,0,0}} -V_{\substack{\boldsymbol{k}_1, \boldsymbol{k}_2, \boldsymbol{k}_4, \boldsymbol{k}\\ 0,0,0,m}}V_{\substack{\boldsymbol{k}, \boldsymbol{k}_3, \boldsymbol{k}_5, \boldsymbol{k}_6\\ m, 0,0,0}}+V_{\substack{\boldsymbol{k}_1, \boldsymbol{k}_2, \boldsymbol{k}_4, \boldsymbol{k}\\ 0,0,0,m}}V_{\substack{\boldsymbol{k}_3, \boldsymbol{k}, \boldsymbol{k}_5, \boldsymbol{k}_6\\ 0, m,0,0}}\right)
\end{equation}
where $V^{\rm{per}}_{\substack{\boldsymbol{k}_1, \boldsymbol{k}_2, \boldsymbol{k}_3, \\ \boldsymbol{k}_4, \boldsymbol{k}_5, \boldsymbol{k}_6}}$ is the interaction strength from the perturbation theory. Expanding each term in the above formula in terms of the form factor, we have:
\begin{enumerate}
    \item 
        $V_{\substack{\boldsymbol{k}_1, \boldsymbol{k}_2, \boldsymbol{k}, \boldsymbol{k}_4\\ 0,0,m,0}}V_{\substack{\boldsymbol{k}, \boldsymbol{k}_3, \boldsymbol{k}_5, \boldsymbol{k}_6\\ m, 0,0,0}}= \sum_{\boldsymbol{g}_1,\boldsymbol{g}_2,\boldsymbol{g}_3,\boldsymbol{g}_4} \delta_{\boldsymbol{k}_1+\boldsymbol{k}_2,\boldsymbol{k}_4+\boldsymbol{k}+\boldsymbol{g}_1+\boldsymbol{g}_2}\delta_{\boldsymbol{k}+\boldsymbol{k}_3,\boldsymbol{k}_5+\boldsymbol{k}_6+\boldsymbol{g}_3+\boldsymbol{g}_4} V(\boldsymbol{k}_1 - \boldsymbol{k}_4 +\boldsymbol{g}_1)V(\boldsymbol{k}_3-\boldsymbol{k}_5+\boldsymbol{g}_4) \times \\F(\boldsymbol{k}_1,\boldsymbol{k}_4,0,0,\boldsymbol{g}_1)F(\boldsymbol{k}_2,\boldsymbol{k},0,m,\boldsymbol{g}_2)F(\boldsymbol{k},\boldsymbol{k}_6,m,0,\boldsymbol{g}_3)F(\boldsymbol{k}_3,\boldsymbol{k}_5,0,0,\boldsymbol{g}_4)$
    \item 
        $V_{\substack{\boldsymbol{k}_1, \boldsymbol{k}_2, \boldsymbol{k}, \boldsymbol{k}_4\\ 0,0,m,0}}V_{\substack{\boldsymbol{k}_3, \boldsymbol{k}, \boldsymbol{k}_5, \boldsymbol{k}_6\\ 0,m,0,0}}= \sum_{\boldsymbol{g}_1,\boldsymbol{g}_2,\boldsymbol{g}_3,\boldsymbol{g}_4} \delta_{\boldsymbol{k}_1+\boldsymbol{k}_2,\boldsymbol{k}_4+\boldsymbol{k}+\boldsymbol{g}_1+\boldsymbol{g}_2}\delta_{\boldsymbol{k}+\boldsymbol{k}_3,\boldsymbol{k}_5+\boldsymbol{k}_6+\boldsymbol{g}_3+\boldsymbol{g}_4} V(\boldsymbol{k}_1 - \boldsymbol{k}_4 +\boldsymbol{g}_1)V(\boldsymbol{k}_3-\boldsymbol{k}_6+\boldsymbol{g}_4) \times \\F(\boldsymbol{k}_1,\boldsymbol{k}_4,0,0,\boldsymbol{g}_1)F(\boldsymbol{k}_2,\boldsymbol{k},0,m,\boldsymbol{g}_2)F(\boldsymbol{k},\boldsymbol{k}_5,m,0,\boldsymbol{g}_3)F(\boldsymbol{k}_3,\boldsymbol{k}_6,0,0,\boldsymbol{g}_4)$
    \item 
        $V_{\substack{\boldsymbol{k}_1, \boldsymbol{k}_2, \boldsymbol{k}_4, \boldsymbol{k}\\ 0,0,0,m}}V_{\substack{\boldsymbol{k}, \boldsymbol{k}_3, \boldsymbol{k}_5, \boldsymbol{k}_6\\ m,0,0,0}}= \sum_{\boldsymbol{g}_1,\boldsymbol{g}_2,\boldsymbol{g}_3,\boldsymbol{g}_4} \delta_{\boldsymbol{k}_1+\boldsymbol{k}_2,\boldsymbol{k}_4+\boldsymbol{k}+\boldsymbol{g}_1+\boldsymbol{g}_2}\delta_{\boldsymbol{k}+\boldsymbol{k}_3,\boldsymbol{k}_5+\boldsymbol{k}_6+\boldsymbol{g}_3+\boldsymbol{g}_4} V(\boldsymbol{k}_2 - \boldsymbol{k}_4 +\boldsymbol{g}_1)V(\boldsymbol{k}_3-\boldsymbol{k}_5+\boldsymbol{g}_4) \times \\F(\boldsymbol{k}_2,\boldsymbol{k}_4,0,0,\boldsymbol{g}_1)F(\boldsymbol{k}_1,\boldsymbol{k},0,m,\boldsymbol{g}_1)F(\boldsymbol{k},\boldsymbol{k}_6,m,0,\boldsymbol{g}_3)F(\boldsymbol{k}_3,\boldsymbol{k}_5,0,0,\boldsymbol{g}_4)$
    \item 
        $V_{\substack{\boldsymbol{k}_1, \boldsymbol{k}_2, \boldsymbol{k}_4, \boldsymbol{k}\\ 0,0,0,m}}V_{\substack{\boldsymbol{k}_3, \boldsymbol{k}, \boldsymbol{k}_5, \boldsymbol{k}_6\\ 0,m,0,0}}= \sum_{\boldsymbol{g}_1,\boldsymbol{g}_2,\boldsymbol{g}_3,\boldsymbol{g}_4} \delta_{\boldsymbol{k}_1+\boldsymbol{k}_2,\boldsymbol{k}_4+\boldsymbol{k}+\boldsymbol{g}_1+\boldsymbol{g}_2}\delta_{\boldsymbol{k}+\boldsymbol{k}_3,\boldsymbol{k}_5+\boldsymbol{k}_6+\boldsymbol{g}_3+\boldsymbol{g}_4} V(\boldsymbol{k}_2 - \boldsymbol{k}_4 +\boldsymbol{g}_2)V(\boldsymbol{k}_3-\boldsymbol{k}_6+\boldsymbol{g}_4) \times \\F(\boldsymbol{k}_2,\boldsymbol{k}_4,0,0,\boldsymbol{g}_1)F(\boldsymbol{k}_1,\boldsymbol{k},0,m,\boldsymbol{g}_2)F(\boldsymbol{k},\boldsymbol{k}_5,m,0,\boldsymbol{g}_3)F(\boldsymbol{k}_3,\boldsymbol{k}_6,0,0,\boldsymbol{g}_4)$
\end{enumerate}

The above terms take the same form instead of permuting $\boldsymbol{k}$ indices. It gives rise to a three-body interaction but it's not obvious this interaction will lead to the Moore-Read Pfaffian ground state since we don't know whether it is short-range. The short-range interaction in main text projected onto the Chern band takes the form:
\begin{equation}
\begin{aligned}
    V_{\substack{\boldsymbol{k}_1, \boldsymbol{k}_2, \boldsymbol{k}_3 \\ \boldsymbol{k}_4, \boldsymbol{k}_5, \boldsymbol{k}_6}}^{\rm 3b} =\sum_{\boldsymbol{g}_1, \boldsymbol{g}_2, \boldsymbol{g}_3}\delta_{\boldsymbol{k}_1+\boldsymbol{k}_2+\boldsymbol{k}_3, \boldsymbol{k}_4+\boldsymbol{k}_5+\boldsymbol{k}_6+\boldsymbol{g}_1+\boldsymbol{g}_2+\boldsymbol{g}_3} V^{\rm 3b}\left(\boldsymbol{k}_1-\boldsymbol{k}_4+\boldsymbol{g}_1,\boldsymbol{k}_3-\boldsymbol{k}_5+\boldsymbol{g}_3 \right)\times \\ F\left(\boldsymbol{k}_1, \boldsymbol{k}_4, 0,0, \boldsymbol{g}_1\right) F\left(\boldsymbol{k}_2, \boldsymbol{k}_6, 0, 0, \boldsymbol{g}_2\right) F\left(\boldsymbol{k}_3, \boldsymbol{k}_5, 0,0, \boldsymbol{g}_3\right)
\end{aligned}
\end{equation}
where $V^{\rm 3b}\left(\boldsymbol{k}_1-\boldsymbol{k}_4+\boldsymbol{g}_1,\boldsymbol{k}_3-\boldsymbol{k}_5+\boldsymbol{g}-3\right)$ is the short-range three-body interaction in the $\boldsymbol{k}$ space. Naively to make the form of the three-body interaction from the perturbation resemble the short-range interaction in the main body of the paper we can naturally contract the middle two form factor $F\left(\boldsymbol{k}_2, \boldsymbol{k}, 0, m, \boldsymbol{g}_2\right) F\left(\boldsymbol{k}, \boldsymbol{k}_6, m, 0, \boldsymbol{g}_3\right)$ to make it a single form factor $F(\boldsymbol{k}_2,\boldsymbol{k}_6,0,0,\boldsymbol{g}_2+\boldsymbol{g}_3)$. Now we assume the form factor $F(\boldsymbol{k}_1,\boldsymbol{k}_2,m,n)$ satisfies the condition \eqref{eq:condition} in the main body. We insert it into the above equations before performing the summation over $m$. The middle two form factors in first equation $F\left(\boldsymbol{k}_2, \boldsymbol{k}, 0, m, \boldsymbol{g}_2\right) F\left(\boldsymbol{k}, \boldsymbol{k}_6, m, 0, \boldsymbol{g}_3\right)$ with energy $E_m$ reduce to one form factor $F(\boldsymbol{k}_2,\boldsymbol{k}_6,0,0,\boldsymbol{g}_2+\boldsymbol{g}_3)$  Then the summation of equation over $\boldsymbol{k},m$ weighted by inverse of energy becomes,
\begin{equation}
\begin{aligned}
\sum_{m,\boldsymbol{k}}\frac{1}{\Delta E_m}V_{\substack{\boldsymbol{k}_1, \boldsymbol{k}_2, \boldsymbol{k}, \boldsymbol{k}_4\\ 0,0,m,0}}V_{\substack{\boldsymbol{k}, \boldsymbol{k}_3, \boldsymbol{k}_5, \boldsymbol{k}_6\\ m, 0,0,0}} &=C \sum_{\boldsymbol{g}_1, \boldsymbol{g}_2, \boldsymbol{g}_3, \boldsymbol{g}_4} \delta_{\boldsymbol{k}_1+\boldsymbol{k}_2 +\boldsymbol{k}_3, \boldsymbol{k}_4+\boldsymbol{k}_5+\boldsymbol{k}_6+\boldsymbol{g}_1+\boldsymbol{g}_2+\boldsymbol{g}_3+\boldsymbol{g}_4} V\left(\boldsymbol{k}_1-\boldsymbol{k}_4+\boldsymbol{g}_1\right) V\left(\boldsymbol{k}_3-\boldsymbol{k}_5+\boldsymbol{g}_4\right) \times \\
& \quad F\left(\boldsymbol{k}_1, \boldsymbol{k}_4, 0,0, \boldsymbol{g}_1\right) F\left(\boldsymbol{k}_2, \boldsymbol{k}_6, 0, 0, \boldsymbol{g}_2+\boldsymbol{g}_3\right)F\left(\boldsymbol{k}_3, \boldsymbol{k}_5, 0,0, \boldsymbol{g}_4\right)
\end{aligned}
\end{equation}
Changing the notation $\boldsymbol{g}_2+\boldsymbol{g}_3 \rightarrow \boldsymbol{g}_2$,\  $\boldsymbol{g}_4 \rightarrow \boldsymbol{g}_3$, the above first equation becomes:
\begin{equation}
\begin{aligned}
\sum_{m,\boldsymbol{k}}\frac{1}{\Delta E_m}V_{\substack{\boldsymbol{k}_1, \boldsymbol{k}_2, \boldsymbol{k}, \boldsymbol{k}_4\\ 0,0,m,0}}V_{\substack{\boldsymbol{k}, \boldsymbol{k}_3, \boldsymbol{k}_5, \boldsymbol{k}_6\\ m, 0,0,0}} &=C \sum_{\boldsymbol{g}_1, \boldsymbol{g}_2, \boldsymbol{g}_3} \delta_{\boldsymbol{k}_1+\boldsymbol{k}_2 +\boldsymbol{k}_3, \boldsymbol{k}_4+\boldsymbol{k}_5+\boldsymbol{k}_6+\boldsymbol{g}_1+\boldsymbol{g}_2+\boldsymbol{g}_3} V\left(\boldsymbol{k}_1-\boldsymbol{k}_4+\boldsymbol{g}_1\right) V\left(\boldsymbol{k}_3-\boldsymbol{k}_5+\boldsymbol{g}_3\right) \times \\
& \quad F\left(\boldsymbol{k}_1, \boldsymbol{k}_4, 0,0, \boldsymbol{g}_1\right) F\left(\boldsymbol{k}_2, \boldsymbol{k}_6, 0, 0, \boldsymbol{g}_2\right)F\left(\boldsymbol{k}_3, \boldsymbol{k}_5, 0,0, \boldsymbol{g}_3\right)
\end{aligned}
\end{equation}
Carry out a similar calculation for the other three and we have the condition of the Pfaffian state to arise
\begin{equation}
\begin{aligned}
    V_{\substack{\boldsymbol{k}_1, \boldsymbol{k}_2, \boldsymbol{k}_3 \\ \boldsymbol{k}_4, \boldsymbol{k}_5, \boldsymbol{k}_6}}^{\rm{per}} =& C \sum_{\boldsymbol{g}_1,\boldsymbol{g}_2,\boldsymbol{g}_3} \delta_{\boldsymbol{k}_1+\boldsymbol{k}_2+\boldsymbol{k}_3,\boldsymbol{k}_4+\boldsymbol{k}_5+\boldsymbol{k}_6+\boldsymbol{g}_1+\boldsymbol{g}_2+\boldsymbol{g}_3} V(\boldsymbol{k}_1 - \boldsymbol{k}_4 +\boldsymbol{g}_1)V(\boldsymbol{k}_3-\boldsymbol{k}_5+\boldsymbol{g}_3) \times \\ &F(\boldsymbol{k}_1,\boldsymbol{k}_4,0,0,\boldsymbol{g}_1)F(\boldsymbol{k}_2,\boldsymbol{k}_6,0,0,\boldsymbol{g}_2)F(\boldsymbol{k}_3,\boldsymbol{k}_5,0,0,\boldsymbol{g}_3)
\end{aligned}
\end{equation}
where the form factors $F(\boldsymbol{k}_1,\boldsymbol{k}_2,0,0,\boldsymbol{g})$ involve the wave function of the single particle Hamiltonian. Because of the statistical properties of fermions, there is an overall factor 4 in front of the interaction incorporated into the constant $C$. Although our condition may not be easily satisfied, it provides a convenient way to determine the resemblance between the three-body interaction from perturbation and the pseudo-potential. Besides, we also expect that if band mixing could soften the Coulomb interaction between the electrons it is more likely the system is in the Moore-Read state. In the thermodynamic limit, we expect the topological ground state should be robust against the non-topological channel. Therefore we can argue the topological ground state will emerge as the dominant state.

%

\appendix

\end{document}

%% file: Main.bbl
%

%% file: Main.bbl
\begin{thebibliography}{53}%
\makeatletter
\providecommand \@ifxundefined [1]{%
 \@ifx{#1\undefined}
}%
\providecommand \@ifnum [1]{%
 \ifnum #1\expandafter \@firstoftwo
 \else \expandafter \@secondoftwo
 \fi
}%
\providecommand \@ifx [1]{%
 \ifx #1\expandafter \@firstoftwo
 \else \expandafter \@secondoftwo
 \fi
}%
\providecommand \natexlab [1]{#1}%
\providecommand \enquote  [1]{``#1''}%
\providecommand \bibnamefont  [1]{#1}%
\providecommand \bibfnamefont [1]{#1}%
\providecommand \citenamefont [1]{#1}%
\providecommand \href@noop [0]{\@secondoftwo}%
\providecommand \href [0]{\begingroup \@sanitize@url \@href}%
\providecommand \@href[1]{\@@startlink{#1}\@@href}%
\providecommand \@@href[1]{\endgroup#1\@@endlink}%
\providecommand \@sanitize@url [0]{\catcode `\\12\catcode `\$12\catcode
  `\&12\catcode `\#12\catcode `\^12\catcode `\_12\catcode `\%12\relax}%
\providecommand \@@startlink[1]{}%
\providecommand \@@endlink[0]{}%
\providecommand \url  [0]{\begingroup\@sanitize@url \@url }%
\providecommand \@url [1]{\endgroup\@href {#1}{\urlprefix }}%
\providecommand \urlprefix  [0]{URL }%
\providecommand \Eprint [0]{\href }%
\providecommand \doibase [0]{http://dx.doi.org/}%
\providecommand \selectlanguage [0]{\@gobble}%
\providecommand \bibinfo  [0]{\@secondoftwo}%
\providecommand \bibfield  [0]{\@secondoftwo}%
\providecommand \translation [1]{[#1]}%
\providecommand \BibitemOpen [0]{}%
\providecommand \bibitemStop [0]{}%
\providecommand \bibitemNoStop [0]{.\EOS\space}%
\providecommand \EOS [0]{\spacefactor3000\relax}%
\providecommand \BibitemShut  [1]{\csname bibitem#1\endcsname}%
\let\auto@bib@innerbib\@empty
\bibitem [{\citenamefont {Moore}\ and\ \citenamefont
  {Read}(1991)}]{MooreNonabelian}%
  \BibitemOpen
  \bibfield  {author} {\bibinfo {author} {\bibfnamefont {G.}~\bibnamefont
  {Moore}}\ and\ \bibinfo {author} {\bibfnamefont {N.}~\bibnamefont {Read}},\
  }\href {\doibase https://doi.org/10.1016/0550-3213(91)90407-O} {\bibfield
  {journal} {\bibinfo  {journal} {Nuclear Physics B}\ }\textbf {\bibinfo
  {volume} {360}},\ \bibinfo {pages} {362} (\bibinfo {year}
  {1991})}\BibitemShut {NoStop}%
\bibitem [{\citenamefont {Wen}(1991)}]{XiaoGangNonabelian}%
  \BibitemOpen
  \bibfield  {author} {\bibinfo {author} {\bibfnamefont {X.~G.}\ \bibnamefont
  {Wen}},\ }\href {\doibase 10.1103/PhysRevLett.66.802} {\bibfield  {journal}
  {\bibinfo  {journal} {Phys. Rev. Lett.}\ }\textbf {\bibinfo {volume} {66}},\
  \bibinfo {pages} {802} (\bibinfo {year} {1991})}\BibitemShut {NoStop}%
\bibitem [{\citenamefont {Greiter}\ \emph {et~al.}(1991)\citenamefont
  {Greiter}, \citenamefont {Wen},\ and\ \citenamefont
  {Wilczek}}]{XiaoGangPaired}%
  \BibitemOpen
  \bibfield  {author} {\bibinfo {author} {\bibfnamefont {M.}~\bibnamefont
  {Greiter}}, \bibinfo {author} {\bibfnamefont {X.-G.}\ \bibnamefont {Wen}}, \
  and\ \bibinfo {author} {\bibfnamefont {F.}~\bibnamefont {Wilczek}},\ }\href
  {\doibase 10.1103/PhysRevLett.66.3205} {\bibfield  {journal} {\bibinfo
  {journal} {Phys. Rev. Lett.}\ }\textbf {\bibinfo {volume} {66}},\ \bibinfo
  {pages} {3205} (\bibinfo {year} {1991})}\BibitemShut {NoStop}%
\bibitem [{\citenamefont {Rezayi}\ and\ \citenamefont
  {Haldane}(2000)}]{RezayiIncompressible}%
  \BibitemOpen
  \bibfield  {author} {\bibinfo {author} {\bibfnamefont {E.~H.}\ \bibnamefont
  {Rezayi}}\ and\ \bibinfo {author} {\bibfnamefont {F.~D.~M.}\ \bibnamefont
  {Haldane}},\ }\href {\doibase 10.1103/PhysRevLett.84.4685} {\bibfield
  {journal} {\bibinfo  {journal} {Phys. Rev. Lett.}\ }\textbf {\bibinfo
  {volume} {84}},\ \bibinfo {pages} {4685} (\bibinfo {year}
  {2000})}\BibitemShut {NoStop}%
\bibitem [{\citenamefont {Read}\ and\ \citenamefont
  {Green}(2000)}]{ReadPaired}%
  \BibitemOpen
  \bibfield  {author} {\bibinfo {author} {\bibfnamefont {N.}~\bibnamefont
  {Read}}\ and\ \bibinfo {author} {\bibfnamefont {D.}~\bibnamefont {Green}},\
  }\href {\doibase 10.1103/PhysRevB.61.10267} {\bibfield  {journal} {\bibinfo
  {journal} {Phys. Rev. B}\ }\textbf {\bibinfo {volume} {61}},\ \bibinfo
  {pages} {10267} (\bibinfo {year} {2000})}\BibitemShut {NoStop}%
\bibitem [{\citenamefont {Li}\ and\ \citenamefont
  {Haldane}(2008)}]{HaldaneEntanglement}%
  \BibitemOpen
  \bibfield  {author} {\bibinfo {author} {\bibfnamefont {H.}~\bibnamefont
  {Li}}\ and\ \bibinfo {author} {\bibfnamefont {F.~D.~M.}\ \bibnamefont
  {Haldane}},\ }\href {\doibase 10.1103/PhysRevLett.101.010504} {\bibfield
  {journal} {\bibinfo  {journal} {Phys. Rev. Lett.}\ }\textbf {\bibinfo
  {volume} {101}},\ \bibinfo {pages} {010504} (\bibinfo {year}
  {2008})}\BibitemShut {NoStop}%
\bibitem [{\citenamefont {Kitaev}(2003)}]{KitaevFault}%
  \BibitemOpen
  \bibfield  {author} {\bibinfo {author} {\bibfnamefont {A.}~\bibnamefont
  {Kitaev}},\ }\href {\doibase https://doi.org/10.1016/S0003-4916(02)00018-0}
  {\bibfield  {journal} {\bibinfo  {journal} {Annals of Physics}\ }\textbf
  {\bibinfo {volume} {303}},\ \bibinfo {pages} {2} (\bibinfo {year}
  {2003})}\BibitemShut {NoStop}%
\bibitem [{\citenamefont {Freedman}\ \emph {et~al.}(2003)\citenamefont
  {Freedman}, \citenamefont {Kitaev}, \citenamefont {Larsen},\ and\
  \citenamefont {Wang}}]{freedman2003topological}%
  \BibitemOpen
  \bibfield  {author} {\bibinfo {author} {\bibfnamefont {M.}~\bibnamefont
  {Freedman}}, \bibinfo {author} {\bibfnamefont {A.}~\bibnamefont {Kitaev}},
  \bibinfo {author} {\bibfnamefont {M.}~\bibnamefont {Larsen}}, \ and\ \bibinfo
  {author} {\bibfnamefont {Z.}~\bibnamefont {Wang}},\ }\href@noop {} {\bibfield
   {journal} {\bibinfo  {journal} {Bulletin of the American Mathematical
  Society}\ }\textbf {\bibinfo {volume} {40}},\ \bibinfo {pages} {31} (\bibinfo
  {year} {2003})}\BibitemShut {NoStop}%
\bibitem [{\citenamefont {Nayak}\ \emph {et~al.}(2008)\citenamefont {Nayak},
  \citenamefont {Simon}, \citenamefont {Stern}, \citenamefont {Freedman},\ and\
  \citenamefont {Das~Sarma}}]{RMPnonabelian}%
  \BibitemOpen
  \bibfield  {author} {\bibinfo {author} {\bibfnamefont {C.}~\bibnamefont
  {Nayak}}, \bibinfo {author} {\bibfnamefont {S.~H.}\ \bibnamefont {Simon}},
  \bibinfo {author} {\bibfnamefont {A.}~\bibnamefont {Stern}}, \bibinfo
  {author} {\bibfnamefont {M.}~\bibnamefont {Freedman}}, \ and\ \bibinfo
  {author} {\bibfnamefont {S.}~\bibnamefont {Das~Sarma}},\ }\href {\doibase
  10.1103/RevModPhys.80.1083} {\bibfield  {journal} {\bibinfo  {journal} {Rev.
  Mod. Phys.}\ }\textbf {\bibinfo {volume} {80}},\ \bibinfo {pages} {1083}
  (\bibinfo {year} {2008})}\BibitemShut {NoStop}%
\bibitem [{\citenamefont {Willett}\ \emph {et~al.}(1987)\citenamefont
  {Willett}, \citenamefont {Eisenstein}, \citenamefont {St{\"o}rmer},
  \citenamefont {Tsui}, \citenamefont {Gossard},\ and\ \citenamefont
  {English}}]{willett1987observation}%
  \BibitemOpen
  \bibfield  {author} {\bibinfo {author} {\bibfnamefont {R.}~\bibnamefont
  {Willett}}, \bibinfo {author} {\bibfnamefont {J.~P.}\ \bibnamefont
  {Eisenstein}}, \bibinfo {author} {\bibfnamefont {H.~L.}\ \bibnamefont
  {St{\"o}rmer}}, \bibinfo {author} {\bibfnamefont {D.~C.}\ \bibnamefont
  {Tsui}}, \bibinfo {author} {\bibfnamefont {A.~C.}\ \bibnamefont {Gossard}}, \
  and\ \bibinfo {author} {\bibfnamefont {J.}~\bibnamefont {English}},\
  }\href@noop {} {\bibfield  {journal} {\bibinfo  {journal} {Physical review
  letters}\ }\textbf {\bibinfo {volume} {59}},\ \bibinfo {pages} {1776}
  (\bibinfo {year} {1987})}\BibitemShut {NoStop}%
\bibitem [{\citenamefont {Storni}\ \emph {et~al.}(2010)\citenamefont {Storni},
  \citenamefont {Morf},\ and\ \citenamefont {Sarma}}]{storni2010fractional}%
  \BibitemOpen
  \bibfield  {author} {\bibinfo {author} {\bibfnamefont {M.}~\bibnamefont
  {Storni}}, \bibinfo {author} {\bibfnamefont {R.}~\bibnamefont {Morf}}, \ and\
  \bibinfo {author} {\bibfnamefont {S.~D.}\ \bibnamefont {Sarma}},\ }\href@noop
  {} {\bibfield  {journal} {\bibinfo  {journal} {Physical review letters}\
  }\textbf {\bibinfo {volume} {104}},\ \bibinfo {pages} {076803} (\bibinfo
  {year} {2010})}\BibitemShut {NoStop}%
\bibitem [{\citenamefont {Bonderson}\ \emph {et~al.}(2006)\citenamefont
  {Bonderson}, \citenamefont {Kitaev},\ and\ \citenamefont
  {Shtengel}}]{bonderson2006detecting}%
  \BibitemOpen
  \bibfield  {author} {\bibinfo {author} {\bibfnamefont {P.}~\bibnamefont
  {Bonderson}}, \bibinfo {author} {\bibfnamefont {A.}~\bibnamefont {Kitaev}}, \
  and\ \bibinfo {author} {\bibfnamefont {K.}~\bibnamefont {Shtengel}},\
  }\href@noop {} {\bibfield  {journal} {\bibinfo  {journal} {Physical review
  letters}\ }\textbf {\bibinfo {volume} {96}},\ \bibinfo {pages} {016803}
  (\bibinfo {year} {2006})}\BibitemShut {NoStop}%
\bibitem [{\citenamefont {W{\'o}js}\ \emph
  {et~al.}(2010{\natexlab{a}})\citenamefont {W{\'o}js}, \citenamefont
  {M{\"o}ller}, \citenamefont {Simon},\ and\ \citenamefont
  {Cooper}}]{wojs2010skyrmions}%
  \BibitemOpen
  \bibfield  {author} {\bibinfo {author} {\bibfnamefont {A.}~\bibnamefont
  {W{\'o}js}}, \bibinfo {author} {\bibfnamefont {G.}~\bibnamefont
  {M{\"o}ller}}, \bibinfo {author} {\bibfnamefont {S.~H.}\ \bibnamefont
  {Simon}}, \ and\ \bibinfo {author} {\bibfnamefont {N.~R.}\ \bibnamefont
  {Cooper}},\ }\href@noop {} {\bibfield  {journal} {\bibinfo  {journal}
  {Physical review letters}\ }\textbf {\bibinfo {volume} {104}},\ \bibinfo
  {pages} {086801} (\bibinfo {year} {2010}{\natexlab{a}})}\BibitemShut
  {NoStop}%
\bibitem [{\citenamefont {Sodemann}\ and\ \citenamefont
  {MacDonald}(2013)}]{MacDonaldmixing}%
  \BibitemOpen
  \bibfield  {author} {\bibinfo {author} {\bibfnamefont {I.}~\bibnamefont
  {Sodemann}}\ and\ \bibinfo {author} {\bibfnamefont {A.~H.}\ \bibnamefont
  {MacDonald}},\ }\href {\doibase 10.1103/PhysRevB.87.245425} {\bibfield
  {journal} {\bibinfo  {journal} {Phys. Rev. B}\ }\textbf {\bibinfo {volume}
  {87}},\ \bibinfo {pages} {245425} (\bibinfo {year} {2013})}\BibitemShut
  {NoStop}%
\bibitem [{\citenamefont {Simon}\ and\ \citenamefont
  {Rezayi}(2013)}]{Rezayimixing}%
  \BibitemOpen
  \bibfield  {author} {\bibinfo {author} {\bibfnamefont {S.~H.}\ \bibnamefont
  {Simon}}\ and\ \bibinfo {author} {\bibfnamefont {E.~H.}\ \bibnamefont
  {Rezayi}},\ }\href
  {https://journals.aps.org/prb/abstract/10.1103/PhysRevB.87.155426} {\bibfield
   {journal} {\bibinfo  {journal} {Physical Review B}\ }\textbf {\bibinfo
  {volume} {87}},\ \bibinfo {pages} {155426} (\bibinfo {year}
  {2013})}\BibitemShut {NoStop}%
\bibitem [{\citenamefont {Rezayi}(2017)}]{Rezayimixing2}%
  \BibitemOpen
  \bibfield  {author} {\bibinfo {author} {\bibfnamefont {E.~H.}\ \bibnamefont
  {Rezayi}},\ }\href
  {https://journals.aps.org/prl/abstract/10.1103/PhysRevLett.119.026801}
  {\bibfield  {journal} {\bibinfo  {journal} {Physical Review Letters}\
  }\textbf {\bibinfo {volume} {119}},\ \bibinfo {pages} {026801} (\bibinfo
  {year} {2017})}\BibitemShut {NoStop}%
\bibitem [{\citenamefont {Peterson}\ \emph {et~al.}(2008)\citenamefont
  {Peterson}, \citenamefont {Park},\ and\ \citenamefont
  {Das~Sarma}}]{Dassarma2008spontaneous}%
  \BibitemOpen
  \bibfield  {author} {\bibinfo {author} {\bibfnamefont {M.~R.}\ \bibnamefont
  {Peterson}}, \bibinfo {author} {\bibfnamefont {K.}~\bibnamefont {Park}}, \
  and\ \bibinfo {author} {\bibfnamefont {S.}~\bibnamefont {Das~Sarma}},\ }\href
  {\doibase 10.1103/PhysRevLett.101.156803} {\bibfield  {journal} {\bibinfo
  {journal} {Phys. Rev. Lett.}\ }\textbf {\bibinfo {volume} {101}},\ \bibinfo
  {pages} {156803} (\bibinfo {year} {2008})}\BibitemShut {NoStop}%
\bibitem [{\citenamefont {Wang}\ \emph {et~al.}(2009)\citenamefont {Wang},
  \citenamefont {Sheng},\ and\ \citenamefont {Haldane}}]{wang2009particle}%
  \BibitemOpen
  \bibfield  {author} {\bibinfo {author} {\bibfnamefont {H.}~\bibnamefont
  {Wang}}, \bibinfo {author} {\bibfnamefont {D.}~\bibnamefont {Sheng}}, \ and\
  \bibinfo {author} {\bibfnamefont {F.}~\bibnamefont {Haldane}},\ }\href@noop
  {} {\bibfield  {journal} {\bibinfo  {journal} {Physical Review B}\ }\textbf
  {\bibinfo {volume} {80}},\ \bibinfo {pages} {241311} (\bibinfo {year}
  {2009})}\BibitemShut {NoStop}%
\bibitem [{\citenamefont {Banerjee}\ \emph {et~al.}(2018)\citenamefont
  {Banerjee}, \citenamefont {Heiblum}, \citenamefont {Umansky}, \citenamefont
  {Feldman}, \citenamefont {Oreg},\ and\ \citenamefont
  {Stern}}]{banerjee2018observation}%
  \BibitemOpen
  \bibfield  {author} {\bibinfo {author} {\bibfnamefont {M.}~\bibnamefont
  {Banerjee}}, \bibinfo {author} {\bibfnamefont {M.}~\bibnamefont {Heiblum}},
  \bibinfo {author} {\bibfnamefont {V.}~\bibnamefont {Umansky}}, \bibinfo
  {author} {\bibfnamefont {D.~E.}\ \bibnamefont {Feldman}}, \bibinfo {author}
  {\bibfnamefont {Y.}~\bibnamefont {Oreg}}, \ and\ \bibinfo {author}
  {\bibfnamefont {A.}~\bibnamefont {Stern}},\ }\href@noop {} {\bibfield
  {journal} {\bibinfo  {journal} {Nature}\ }\textbf {\bibinfo {volume} {559}},\
  \bibinfo {pages} {205} (\bibinfo {year} {2018})}\BibitemShut {NoStop}%
\bibitem [{\citenamefont {Zibrov}\ \emph {et~al.}(2018)\citenamefont {Zibrov},
  \citenamefont {Spanton}, \citenamefont {Zhou}, \citenamefont {Kometter},
  \citenamefont {Taniguchi}, \citenamefont {Watanabe},\ and\ \citenamefont
  {Young}}]{zibrov2018even}%
  \BibitemOpen
  \bibfield  {author} {\bibinfo {author} {\bibfnamefont {A.~A.}\ \bibnamefont
  {Zibrov}}, \bibinfo {author} {\bibfnamefont {E.~M.}\ \bibnamefont {Spanton}},
  \bibinfo {author} {\bibfnamefont {H.}~\bibnamefont {Zhou}}, \bibinfo {author}
  {\bibfnamefont {C.}~\bibnamefont {Kometter}}, \bibinfo {author}
  {\bibfnamefont {T.}~\bibnamefont {Taniguchi}}, \bibinfo {author}
  {\bibfnamefont {K.}~\bibnamefont {Watanabe}}, \ and\ \bibinfo {author}
  {\bibfnamefont {A.~F.}\ \bibnamefont {Young}},\ }\href {\doibase
  10.1038/s41567-018-0190-0} {\bibfield  {journal} {\bibinfo  {journal} {Nature
  Physics}\ }\textbf {\bibinfo {volume} {14}},\ \bibinfo {pages} {930}
  (\bibinfo {year} {2018})}\BibitemShut {NoStop}%
\bibitem [{\citenamefont {Kang}\ \emph {et~al.}(2024)\citenamefont {Kang},
  \citenamefont {Shen}, \citenamefont {Qiu}, \citenamefont {Watanabe},
  \citenamefont {Taniguchi}, \citenamefont {Shan},\ and\ \citenamefont
  {Mak}}]{kang2024observation}%
  \BibitemOpen
  \bibfield  {author} {\bibinfo {author} {\bibfnamefont {K.}~\bibnamefont
  {Kang}}, \bibinfo {author} {\bibfnamefont {B.}~\bibnamefont {Shen}}, \bibinfo
  {author} {\bibfnamefont {Y.}~\bibnamefont {Qiu}}, \bibinfo {author}
  {\bibfnamefont {K.}~\bibnamefont {Watanabe}}, \bibinfo {author}
  {\bibfnamefont {T.}~\bibnamefont {Taniguchi}}, \bibinfo {author}
  {\bibfnamefont {J.}~\bibnamefont {Shan}}, \ and\ \bibinfo {author}
  {\bibfnamefont {K.~F.}\ \bibnamefont {Mak}},\ }\href@noop {} {\bibfield
  {journal} {\bibinfo  {journal} {arXiv preprint arXiv:2402.03294}\ } (\bibinfo
  {year} {2024})}\BibitemShut {NoStop}%
\bibitem [{\citenamefont {Zhang}\ \emph {et~al.}(2023)\citenamefont {Zhang},
  \citenamefont {Wang}, \citenamefont {Mishra},\ and\ \citenamefont
  {Liu}}]{Zhang2023TwoDimensional}%
  \BibitemOpen
  \bibfield  {author} {\bibinfo {author} {\bibfnamefont {C.}~\bibnamefont
  {Zhang}}, \bibinfo {author} {\bibfnamefont {R.}~\bibnamefont {Wang}},
  \bibinfo {author} {\bibfnamefont {H.}~\bibnamefont {Mishra}}, \ and\ \bibinfo
  {author} {\bibfnamefont {Y.}~\bibnamefont {Liu}},\ }\href {\doibase
  10.1103/PhysRevLett.130.087001} {\bibfield  {journal} {\bibinfo  {journal}
  {Phys. Rev. Lett.}\ }\textbf {\bibinfo {volume} {130}},\ \bibinfo {pages}
  {087001} (\bibinfo {year} {2023})}\BibitemShut {NoStop}%
\bibitem [{\citenamefont {Cai}\ \emph {et~al.}(2023)\citenamefont {Cai},
  \citenamefont {Anderson}, \citenamefont {Wang}, \citenamefont {Zhang},
  \citenamefont {Liu}, \citenamefont {Holtzmann}, \citenamefont {Zhang},
  \citenamefont {Fan}, \citenamefont {Taniguchi}, \citenamefont {Watanabe},
  \citenamefont {Ran}, \citenamefont {Cao}, \citenamefont {Fu}, \citenamefont
  {Xiao}, \citenamefont {Yao},\ and\ \citenamefont {Xu}}]{JiaqiFQAH}%
  \BibitemOpen
  \bibfield  {author} {\bibinfo {author} {\bibfnamefont {J.}~\bibnamefont
  {Cai}}, \bibinfo {author} {\bibfnamefont {E.}~\bibnamefont {Anderson}},
  \bibinfo {author} {\bibfnamefont {C.}~\bibnamefont {Wang}}, \bibinfo {author}
  {\bibfnamefont {X.}~\bibnamefont {Zhang}}, \bibinfo {author} {\bibfnamefont
  {X.}~\bibnamefont {Liu}}, \bibinfo {author} {\bibfnamefont {W.}~\bibnamefont
  {Holtzmann}}, \bibinfo {author} {\bibfnamefont {Y.}~\bibnamefont {Zhang}},
  \bibinfo {author} {\bibfnamefont {F.}~\bibnamefont {Fan}}, \bibinfo {author}
  {\bibfnamefont {T.}~\bibnamefont {Taniguchi}}, \bibinfo {author}
  {\bibfnamefont {K.}~\bibnamefont {Watanabe}}, \bibinfo {author}
  {\bibfnamefont {Y.}~\bibnamefont {Ran}}, \bibinfo {author} {\bibfnamefont
  {T.}~\bibnamefont {Cao}}, \bibinfo {author} {\bibfnamefont {L.}~\bibnamefont
  {Fu}}, \bibinfo {author} {\bibfnamefont {D.}~\bibnamefont {Xiao}}, \bibinfo
  {author} {\bibfnamefont {W.}~\bibnamefont {Yao}}, \ and\ \bibinfo {author}
  {\bibfnamefont {X.}~\bibnamefont {Xu}},\ }\href {\doibase
  10.1038/s41586-023-06289-w} {\bibfield  {journal} {\bibinfo  {journal}
  {Nature}\ }\textbf {\bibinfo {volume} {622}},\ \bibinfo {pages} {63}
  (\bibinfo {year} {2023})}\BibitemShut {NoStop}%
\bibitem [{\citenamefont {Lu}\ \emph {et~al.}(2023)\citenamefont {Lu},
  \citenamefont {Han}, \citenamefont {Yao}, \citenamefont {Reddy},
  \citenamefont {Yang}, \citenamefont {Seo}, \citenamefont {Watanabe},
  \citenamefont {Taniguchi}, \citenamefont {Fu},\ and\ \citenamefont
  {Ju}}]{ZhengguangFractional}%
  \BibitemOpen
  \bibfield  {author} {\bibinfo {author} {\bibfnamefont {Z.}~\bibnamefont
  {Lu}}, \bibinfo {author} {\bibfnamefont {T.}~\bibnamefont {Han}}, \bibinfo
  {author} {\bibfnamefont {Y.}~\bibnamefont {Yao}}, \bibinfo {author}
  {\bibfnamefont {A.~P.}\ \bibnamefont {Reddy}}, \bibinfo {author}
  {\bibfnamefont {J.}~\bibnamefont {Yang}}, \bibinfo {author} {\bibfnamefont
  {J.}~\bibnamefont {Seo}}, \bibinfo {author} {\bibfnamefont {K.}~\bibnamefont
  {Watanabe}}, \bibinfo {author} {\bibfnamefont {T.}~\bibnamefont {Taniguchi}},
  \bibinfo {author} {\bibfnamefont {L.}~\bibnamefont {Fu}}, \ and\ \bibinfo
  {author} {\bibfnamefont {L.}~\bibnamefont {Ju}},\ }\href@noop {} {\enquote
  {\bibinfo {title} {{Fractional Quantum Anomalous Hall Effect in a Graphene
  Moire Superlattice}},}\ } (\bibinfo {year} {2023}),\ \Eprint
  {http://arxiv.org/abs/2309.17436} {arXiv:2309.17436 [cond-mat.mes-hall]}
  \BibitemShut {NoStop}%
\bibitem [{\citenamefont {Roy}(2014)}]{roy2014band}%
  \BibitemOpen
  \bibfield  {author} {\bibinfo {author} {\bibfnamefont {R.}~\bibnamefont
  {Roy}},\ }\href@noop {} {\bibfield  {journal} {\bibinfo  {journal} {Physical
  Review B}\ }\textbf {\bibinfo {volume} {90}},\ \bibinfo {pages} {165139}
  (\bibinfo {year} {2014})}\BibitemShut {NoStop}%
\bibitem [{\citenamefont {Reddy}\ \emph {et~al.}(2023)\citenamefont {Reddy},
  \citenamefont {Alsallom}, \citenamefont {Zhang}, \citenamefont {Devakul},\
  and\ \citenamefont {Fu}}]{LiangFuFQAH}%
  \BibitemOpen
  \bibfield  {author} {\bibinfo {author} {\bibfnamefont {A.~P.}\ \bibnamefont
  {Reddy}}, \bibinfo {author} {\bibfnamefont {F.}~\bibnamefont {Alsallom}},
  \bibinfo {author} {\bibfnamefont {Y.}~\bibnamefont {Zhang}}, \bibinfo
  {author} {\bibfnamefont {T.}~\bibnamefont {Devakul}}, \ and\ \bibinfo
  {author} {\bibfnamefont {L.}~\bibnamefont {Fu}},\ }\href {\doibase
  10.1103/PhysRevB.108.085117} {\bibfield  {journal} {\bibinfo  {journal}
  {Phys. Rev. B}\ }\textbf {\bibinfo {volume} {108}},\ \bibinfo {pages}
  {085117} (\bibinfo {year} {2023})}\BibitemShut {NoStop}%
\bibitem [{\citenamefont {Dong}\ \emph
  {et~al.}(2023{\natexlab{a}})\citenamefont {Dong}, \citenamefont {Wang},
  \citenamefont {Ledwith}, \citenamefont {Vishwanath},\ and\ \citenamefont
  {Parker}}]{AshvinCFL}%
  \BibitemOpen
  \bibfield  {author} {\bibinfo {author} {\bibfnamefont {J.}~\bibnamefont
  {Dong}}, \bibinfo {author} {\bibfnamefont {J.}~\bibnamefont {Wang}}, \bibinfo
  {author} {\bibfnamefont {P.~J.}\ \bibnamefont {Ledwith}}, \bibinfo {author}
  {\bibfnamefont {A.}~\bibnamefont {Vishwanath}}, \ and\ \bibinfo {author}
  {\bibfnamefont {D.~E.}\ \bibnamefont {Parker}},\ }\href {\doibase
  10.1103/PhysRevLett.131.136502} {\bibfield  {journal} {\bibinfo  {journal}
  {Phys. Rev. Lett.}\ }\textbf {\bibinfo {volume} {131}},\ \bibinfo {pages}
  {136502} (\bibinfo {year} {2023}{\natexlab{a}})}\BibitemShut {NoStop}%
\bibitem [{\citenamefont {Goldman}\ \emph {et~al.}(2023)\citenamefont
  {Goldman}, \citenamefont {Reddy}, \citenamefont {Paul},\ and\ \citenamefont
  {Fu}}]{LiangFuCFL}%
  \BibitemOpen
  \bibfield  {author} {\bibinfo {author} {\bibfnamefont {H.}~\bibnamefont
  {Goldman}}, \bibinfo {author} {\bibfnamefont {A.~P.}\ \bibnamefont {Reddy}},
  \bibinfo {author} {\bibfnamefont {N.}~\bibnamefont {Paul}}, \ and\ \bibinfo
  {author} {\bibfnamefont {L.}~\bibnamefont {Fu}},\ }\href {\doibase
  10.1103/PhysRevLett.131.136501} {\bibfield  {journal} {\bibinfo  {journal}
  {Phys. Rev. Lett.}\ }\textbf {\bibinfo {volume} {131}},\ \bibinfo {pages}
  {136501} (\bibinfo {year} {2023})}\BibitemShut {NoStop}%
\bibitem [{\citenamefont {Wang}\ \emph {et~al.}(2024)\citenamefont {Wang},
  \citenamefont {Zhang}, \citenamefont {Liu}, \citenamefont {He}, \citenamefont
  {Xu}, \citenamefont {Ran}, \citenamefont {Cao},\ and\ \citenamefont
  {Xiao}}]{XiaodiFractional}%
  \BibitemOpen
  \bibfield  {author} {\bibinfo {author} {\bibfnamefont {C.}~\bibnamefont
  {Wang}}, \bibinfo {author} {\bibfnamefont {X.-W.}\ \bibnamefont {Zhang}},
  \bibinfo {author} {\bibfnamefont {X.}~\bibnamefont {Liu}}, \bibinfo {author}
  {\bibfnamefont {Y.}~\bibnamefont {He}}, \bibinfo {author} {\bibfnamefont
  {X.}~\bibnamefont {Xu}}, \bibinfo {author} {\bibfnamefont {Y.}~\bibnamefont
  {Ran}}, \bibinfo {author} {\bibfnamefont {T.}~\bibnamefont {Cao}}, \ and\
  \bibinfo {author} {\bibfnamefont {D.}~\bibnamefont {Xiao}},\ }\href {\doibase
  10.1103/PhysRevLett.132.036501} {\bibfield  {journal} {\bibinfo  {journal}
  {Phys. Rev. Lett.}\ }\textbf {\bibinfo {volume} {132}},\ \bibinfo {pages}
  {036501} (\bibinfo {year} {2024})}\BibitemShut {NoStop}%
\bibitem [{\citenamefont {Song}\ \emph {et~al.}(2024)\citenamefont {Song},
  \citenamefont {Zhang},\ and\ \citenamefont {Senthil}}]{song2024phase}%
  \BibitemOpen
  \bibfield  {author} {\bibinfo {author} {\bibfnamefont {X.-Y.}\ \bibnamefont
  {Song}}, \bibinfo {author} {\bibfnamefont {Y.-H.}\ \bibnamefont {Zhang}}, \
  and\ \bibinfo {author} {\bibfnamefont {T.}~\bibnamefont {Senthil}},\
  }\href@noop {} {\bibfield  {journal} {\bibinfo  {journal} {Physical Review
  B}\ }\textbf {\bibinfo {volume} {109}},\ \bibinfo {pages} {085143} (\bibinfo
  {year} {2024})}\BibitemShut {NoStop}%
\bibitem [{\citenamefont {Devakul}\ \emph {et~al.}(2021)\citenamefont
  {Devakul}, \citenamefont {Cr{\'e}pel}, \citenamefont {Zhang},\ and\
  \citenamefont {Fu}}]{devakul2021magic}%
  \BibitemOpen
  \bibfield  {author} {\bibinfo {author} {\bibfnamefont {T.}~\bibnamefont
  {Devakul}}, \bibinfo {author} {\bibfnamefont {V.}~\bibnamefont {Cr{\'e}pel}},
  \bibinfo {author} {\bibfnamefont {Y.}~\bibnamefont {Zhang}}, \ and\ \bibinfo
  {author} {\bibfnamefont {L.}~\bibnamefont {Fu}},\ }\href@noop {} {\bibfield
  {journal} {\bibinfo  {journal} {Nature communications}\ }\textbf {\bibinfo
  {volume} {12}},\ \bibinfo {pages} {6730} (\bibinfo {year}
  {2021})}\BibitemShut {NoStop}%
\bibitem [{\citenamefont {Bonesteel}(1999)}]{bonestell1999singular}%
  \BibitemOpen
  \bibfield  {author} {\bibinfo {author} {\bibfnamefont {N.~E.}\ \bibnamefont
  {Bonesteel}},\ }\href {\doibase 10.1103/PhysRevLett.82.984} {\bibfield
  {journal} {\bibinfo  {journal} {Phys. Rev. Lett.}\ }\textbf {\bibinfo
  {volume} {82}},\ \bibinfo {pages} {984} (\bibinfo {year} {1999})}\BibitemShut
  {NoStop}%
\bibitem [{\citenamefont {Metlitski}\ \emph {et~al.}(2015)\citenamefont
  {Metlitski}, \citenamefont {Mross}, \citenamefont {Sachdev},\ and\
  \citenamefont {Senthil}}]{metlitski2015cooper}%
  \BibitemOpen
  \bibfield  {author} {\bibinfo {author} {\bibfnamefont {M.~A.}\ \bibnamefont
  {Metlitski}}, \bibinfo {author} {\bibfnamefont {D.~F.}\ \bibnamefont
  {Mross}}, \bibinfo {author} {\bibfnamefont {S.}~\bibnamefont {Sachdev}}, \
  and\ \bibinfo {author} {\bibfnamefont {T.}~\bibnamefont {Senthil}},\ }\href
  {\doibase 10.1103/PhysRevB.91.115111} {\bibfield  {journal} {\bibinfo
  {journal} {Phys. Rev. B}\ }\textbf {\bibinfo {volume} {91}},\ \bibinfo
  {pages} {115111} (\bibinfo {year} {2015})}\BibitemShut {NoStop}%
\bibitem [{\citenamefont {Yang}(2018)}]{BoThreebody}%
  \BibitemOpen
  \bibfield  {author} {\bibinfo {author} {\bibfnamefont {B.}~\bibnamefont
  {Yang}},\ }\href {\doibase 10.1103/PhysRevB.98.201101} {\bibfield  {journal}
  {\bibinfo  {journal} {Phys. Rev. B}\ }\textbf {\bibinfo {volume} {98}},\
  \bibinfo {pages} {201101} (\bibinfo {year} {2018})}\BibitemShut {NoStop}%
\bibitem [{\citenamefont {Xu}\ \emph {et~al.}(2024)\citenamefont {Xu},
  \citenamefont {Li}, \citenamefont {Xu}, \citenamefont {Bi},\ and\
  \citenamefont {Zhang}}]{YangMix}%
  \BibitemOpen
  \bibfield  {author} {\bibinfo {author} {\bibfnamefont {C.}~\bibnamefont
  {Xu}}, \bibinfo {author} {\bibfnamefont {J.}~\bibnamefont {Li}}, \bibinfo
  {author} {\bibfnamefont {Y.}~\bibnamefont {Xu}}, \bibinfo {author}
  {\bibfnamefont {Z.}~\bibnamefont {Bi}}, \ and\ \bibinfo {author}
  {\bibfnamefont {Y.}~\bibnamefont {Zhang}},\ }\href@noop {} {\enquote
  {\bibinfo {title} {{Maximally Localized Wannier Orbitals, Interaction Models
  and Fractional Quantum Anomalous Hall Effect in Twisted Bilayer MoTe2}},}\ }
  (\bibinfo {year} {2024}),\ \Eprint {http://arxiv.org/abs/2308.09697}
  {arXiv:2308.09697 [cond-mat.str-el]} \BibitemShut {NoStop}%
\bibitem [{\citenamefont {W{\'o}js}\ \emph
  {et~al.}(2010{\natexlab{b}})\citenamefont {W{\'o}js}, \citenamefont
  {T{\H{o}}ke},\ and\ \citenamefont {Jain}}]{wojs2010global}%
  \BibitemOpen
  \bibfield  {author} {\bibinfo {author} {\bibfnamefont {A.}~\bibnamefont
  {W{\'o}js}}, \bibinfo {author} {\bibfnamefont {C.}~\bibnamefont
  {T{\H{o}}ke}}, \ and\ \bibinfo {author} {\bibfnamefont {J.~K.}\ \bibnamefont
  {Jain}},\ }\href@noop {} {\bibfield  {journal} {\bibinfo  {journal} {Physical
  review letters}\ }\textbf {\bibinfo {volume} {105}},\ \bibinfo {pages}
  {196801} (\bibinfo {year} {2010}{\natexlab{b}})}\BibitemShut {NoStop}%
\bibitem [{\citenamefont {Wu}\ \emph {et~al.}(2012)\citenamefont {Wu},
  \citenamefont {Bernevig},\ and\ \citenamefont {Regnault}}]{BernevigZoology}%
  \BibitemOpen
  \bibfield  {author} {\bibinfo {author} {\bibfnamefont {Y.-L.}\ \bibnamefont
  {Wu}}, \bibinfo {author} {\bibfnamefont {B.~A.}\ \bibnamefont {Bernevig}}, \
  and\ \bibinfo {author} {\bibfnamefont {N.}~\bibnamefont {Regnault}},\ }\href
  {\doibase 10.1103/PhysRevB.85.075116} {\bibfield  {journal} {\bibinfo
  {journal} {Phys. Rev. B}\ }\textbf {\bibinfo {volume} {85}},\ \bibinfo
  {pages} {075116} (\bibinfo {year} {2012})}\BibitemShut {NoStop}%
\bibitem [{\citenamefont {Wang}\ \emph {et~al.}(2012)\citenamefont {Wang},
  \citenamefont {Yao}, \citenamefont {Gu}, \citenamefont {Gong},\ and\
  \citenamefont {Sheng}}]{ShengNonabelian}%
  \BibitemOpen
  \bibfield  {author} {\bibinfo {author} {\bibfnamefont {Y.-F.}\ \bibnamefont
  {Wang}}, \bibinfo {author} {\bibfnamefont {H.}~\bibnamefont {Yao}}, \bibinfo
  {author} {\bibfnamefont {Z.-C.}\ \bibnamefont {Gu}}, \bibinfo {author}
  {\bibfnamefont {C.-D.}\ \bibnamefont {Gong}}, \ and\ \bibinfo {author}
  {\bibfnamefont {D.~N.}\ \bibnamefont {Sheng}},\ }\href {\doibase
  10.1103/PhysRevLett.108.126805} {\bibfield  {journal} {\bibinfo  {journal}
  {Phys. Rev. Lett.}\ }\textbf {\bibinfo {volume} {108}},\ \bibinfo {pages}
  {126805} (\bibinfo {year} {2012})}\BibitemShut {NoStop}%
\bibitem [{\citenamefont {Bernevig}\ and\ \citenamefont
  {Regnault}(2012)}]{BernevigTranslation}%
  \BibitemOpen
  \bibfield  {author} {\bibinfo {author} {\bibfnamefont {B.~A.}\ \bibnamefont
  {Bernevig}}\ and\ \bibinfo {author} {\bibfnamefont {N.}~\bibnamefont
  {Regnault}},\ }\href {\doibase 10.1103/PhysRevB.85.075128} {\bibfield
  {journal} {\bibinfo  {journal} {Phys. Rev. B}\ }\textbf {\bibinfo {volume}
  {85}},\ \bibinfo {pages} {075128} (\bibinfo {year} {2012})}\BibitemShut
  {NoStop}%
\bibitem [{\citenamefont {Wu}\ \emph {et~al.}(2019)\citenamefont {Wu},
  \citenamefont {Lovorn}, \citenamefont {Tutuc}, \citenamefont {Martin},\ and\
  \citenamefont {MacDonald}}]{wu2019topological}%
  \BibitemOpen
  \bibfield  {author} {\bibinfo {author} {\bibfnamefont {F.}~\bibnamefont
  {Wu}}, \bibinfo {author} {\bibfnamefont {T.}~\bibnamefont {Lovorn}}, \bibinfo
  {author} {\bibfnamefont {E.}~\bibnamefont {Tutuc}}, \bibinfo {author}
  {\bibfnamefont {I.}~\bibnamefont {Martin}}, \ and\ \bibinfo {author}
  {\bibfnamefont {A.}~\bibnamefont {MacDonald}},\ }\href@noop {} {\bibfield
  {journal} {\bibinfo  {journal} {Physical review letters}\ }\textbf {\bibinfo
  {volume} {122}},\ \bibinfo {pages} {086402} (\bibinfo {year}
  {2019})}\BibitemShut {NoStop}%
\bibitem [{\citenamefont {Li}\ \emph {et~al.}(2021)\citenamefont {Li},
  \citenamefont {Kumar}, \citenamefont {Sun},\ and\ \citenamefont
  {Lin}}]{li2021spontaneous}%
  \BibitemOpen
  \bibfield  {author} {\bibinfo {author} {\bibfnamefont {H.}~\bibnamefont
  {Li}}, \bibinfo {author} {\bibfnamefont {U.}~\bibnamefont {Kumar}}, \bibinfo
  {author} {\bibfnamefont {K.}~\bibnamefont {Sun}}, \ and\ \bibinfo {author}
  {\bibfnamefont {S.-Z.}\ \bibnamefont {Lin}},\ }\href@noop {} {\bibfield
  {journal} {\bibinfo  {journal} {Physical Review Research}\ }\textbf {\bibinfo
  {volume} {3}},\ \bibinfo {pages} {L032070} (\bibinfo {year}
  {2021})}\BibitemShut {NoStop}%
\bibitem [{\citenamefont {Cr\'epel}\ and\ \citenamefont
  {Fu}(2023)}]{crepel2023fci}%
  \BibitemOpen
  \bibfield  {author} {\bibinfo {author} {\bibfnamefont {V.}~\bibnamefont
  {Cr\'epel}}\ and\ \bibinfo {author} {\bibfnamefont {L.}~\bibnamefont {Fu}},\
  }\href {\doibase 10.1103/PhysRevB.107.L201109} {\bibfield  {journal}
  {\bibinfo  {journal} {Phys. Rev. B}\ }\textbf {\bibinfo {volume} {107}},\
  \bibinfo {pages} {L201109} (\bibinfo {year} {2023})}\BibitemShut {NoStop}%
\bibitem [{\citenamefont {Wang}\ \emph {et~al.}(2021)\citenamefont {Wang},
  \citenamefont {Cano}, \citenamefont {Millis}, \citenamefont {Liu},\ and\
  \citenamefont {Yang}}]{wang2021exact}%
  \BibitemOpen
  \bibfield  {author} {\bibinfo {author} {\bibfnamefont {J.}~\bibnamefont
  {Wang}}, \bibinfo {author} {\bibfnamefont {J.}~\bibnamefont {Cano}}, \bibinfo
  {author} {\bibfnamefont {A.~J.}\ \bibnamefont {Millis}}, \bibinfo {author}
  {\bibfnamefont {Z.}~\bibnamefont {Liu}}, \ and\ \bibinfo {author}
  {\bibfnamefont {B.}~\bibnamefont {Yang}},\ }\href@noop {} {\bibfield
  {journal} {\bibinfo  {journal} {Physical review letters}\ }\textbf {\bibinfo
  {volume} {127}},\ \bibinfo {pages} {246403} (\bibinfo {year}
  {2021})}\BibitemShut {NoStop}%
\bibitem [{\citenamefont {Ledwith}\ \emph {et~al.}(2023)\citenamefont
  {Ledwith}, \citenamefont {Vishwanath},\ and\ \citenamefont
  {Parker}}]{ledwith2023vortexability}%
  \BibitemOpen
  \bibfield  {author} {\bibinfo {author} {\bibfnamefont {P.~J.}\ \bibnamefont
  {Ledwith}}, \bibinfo {author} {\bibfnamefont {A.}~\bibnamefont {Vishwanath}},
  \ and\ \bibinfo {author} {\bibfnamefont {D.~E.}\ \bibnamefont {Parker}},\
  }\href@noop {} {\bibfield  {journal} {\bibinfo  {journal} {Physical Review
  B}\ }\textbf {\bibinfo {volume} {108}},\ \bibinfo {pages} {205144} (\bibinfo
  {year} {2023})}\BibitemShut {NoStop}%
\bibitem [{\citenamefont {Ozawa}\ and\ \citenamefont
  {Mera}(2021)}]{ozawa2021relations}%
  \BibitemOpen
  \bibfield  {author} {\bibinfo {author} {\bibfnamefont {T.}~\bibnamefont
  {Ozawa}}\ and\ \bibinfo {author} {\bibfnamefont {B.}~\bibnamefont {Mera}},\
  }\href@noop {} {\bibfield  {journal} {\bibinfo  {journal} {Physical Review
  B}\ }\textbf {\bibinfo {volume} {104}},\ \bibinfo {pages} {045103} (\bibinfo
  {year} {2021})}\BibitemShut {NoStop}%
\bibitem [{\citenamefont {Oshikawa}\ \emph {et~al.}(2007)\citenamefont
  {Oshikawa}, \citenamefont {Kim}, \citenamefont {Shtengel}, \citenamefont
  {Nayak},\ and\ \citenamefont {Tewari}}]{oshikawa2007topological}%
  \BibitemOpen
  \bibfield  {author} {\bibinfo {author} {\bibfnamefont {M.}~\bibnamefont
  {Oshikawa}}, \bibinfo {author} {\bibfnamefont {Y.~B.}\ \bibnamefont {Kim}},
  \bibinfo {author} {\bibfnamefont {K.}~\bibnamefont {Shtengel}}, \bibinfo
  {author} {\bibfnamefont {C.}~\bibnamefont {Nayak}}, \ and\ \bibinfo {author}
  {\bibfnamefont {S.}~\bibnamefont {Tewari}},\ }\href@noop {} {\bibfield
  {journal} {\bibinfo  {journal} {Annals of Physics}\ }\textbf {\bibinfo
  {volume} {322}},\ \bibinfo {pages} {1477} (\bibinfo {year}
  {2007})}\BibitemShut {NoStop}%
\bibitem [{\citenamefont {Regnault}\ and\ \citenamefont
  {Bernevig}(2011)}]{BernevigFCI}%
  \BibitemOpen
  \bibfield  {author} {\bibinfo {author} {\bibfnamefont {N.}~\bibnamefont
  {Regnault}}\ and\ \bibinfo {author} {\bibfnamefont {B.~A.}\ \bibnamefont
  {Bernevig}},\ }\href@noop {} {\bibfield  {journal} {\bibinfo  {journal}
  {Physical Review X}\ }\textbf {\bibinfo {volume} {1}},\ \bibinfo {pages}
  {021014} (\bibinfo {year} {2011})}\BibitemShut {NoStop}%
\bibitem [{\citenamefont {Wang}\ \emph {et~al.}(2014)\citenamefont {Wang},
  \citenamefont {Mandal}, \citenamefont {Chung},\ and\ \citenamefont
  {Chakravarty}}]{WANG2014pairing}%
  \BibitemOpen
  \bibfield  {author} {\bibinfo {author} {\bibfnamefont {Z.}~\bibnamefont
  {Wang}}, \bibinfo {author} {\bibfnamefont {I.}~\bibnamefont {Mandal}},
  \bibinfo {author} {\bibfnamefont {S.~B.}\ \bibnamefont {Chung}}, \ and\
  \bibinfo {author} {\bibfnamefont {S.}~\bibnamefont {Chakravarty}},\ }\href
  {\doibase https://doi.org/10.1016/j.aop.2014.09.021} {\bibfield  {journal}
  {\bibinfo  {journal} {Annals of Physics}\ }\textbf {\bibinfo {volume}
  {351}},\ \bibinfo {pages} {727} (\bibinfo {year} {2014})}\BibitemShut
  {NoStop}%
\bibitem [{\citenamefont {Coleman}(2015)}]{coleman2015introduction}%
  \BibitemOpen
  \bibfield  {author} {\bibinfo {author} {\bibfnamefont {P.}~\bibnamefont
  {Coleman}},\ }\href@noop {} {\emph {\bibinfo {title} {{Introduction to
  many-body physics}}}}\ (\bibinfo  {publisher} {Cambridge University Press},\
  \bibinfo {year} {2015})\BibitemShut {NoStop}%
\bibitem [{\citenamefont {Gao}\ \emph {et~al.}(2023)\citenamefont {Gao},
  \citenamefont {Dong}, \citenamefont {Ledwith}, \citenamefont {Parker},\ and\
  \citenamefont {Khalaf}}]{gao2023untwisting}%
  \BibitemOpen
  \bibfield  {author} {\bibinfo {author} {\bibfnamefont {Q.}~\bibnamefont
  {Gao}}, \bibinfo {author} {\bibfnamefont {J.}~\bibnamefont {Dong}}, \bibinfo
  {author} {\bibfnamefont {P.}~\bibnamefont {Ledwith}}, \bibinfo {author}
  {\bibfnamefont {D.}~\bibnamefont {Parker}}, \ and\ \bibinfo {author}
  {\bibfnamefont {E.}~\bibnamefont {Khalaf}},\ }\href@noop {} {\bibfield
  {journal} {\bibinfo  {journal} {Physical Review Letters}\ }\textbf {\bibinfo
  {volume} {131}},\ \bibinfo {pages} {096401} (\bibinfo {year}
  {2023})}\BibitemShut {NoStop}%
\bibitem [{\citenamefont {Ledwith}\ \emph {et~al.}(2022)\citenamefont
  {Ledwith}, \citenamefont {Vishwanath},\ and\ \citenamefont
  {Khalaf}}]{ledwith2022family}%
  \BibitemOpen
  \bibfield  {author} {\bibinfo {author} {\bibfnamefont {P.~J.}\ \bibnamefont
  {Ledwith}}, \bibinfo {author} {\bibfnamefont {A.}~\bibnamefont {Vishwanath}},
  \ and\ \bibinfo {author} {\bibfnamefont {E.}~\bibnamefont {Khalaf}},\
  }\href@noop {} {\bibfield  {journal} {\bibinfo  {journal} {Physical Review
  Letters}\ }\textbf {\bibinfo {volume} {128}},\ \bibinfo {pages} {176404}
  (\bibinfo {year} {2022})}\BibitemShut {NoStop}%
\bibitem [{\citenamefont {Ledwith}\ \emph {et~al.}(2020)\citenamefont
  {Ledwith}, \citenamefont {Tarnopolsky}, \citenamefont {Khalaf},\ and\
  \citenamefont {Vishwanath}}]{ledwith2020fractional}%
  \BibitemOpen
  \bibfield  {author} {\bibinfo {author} {\bibfnamefont {P.~J.}\ \bibnamefont
  {Ledwith}}, \bibinfo {author} {\bibfnamefont {G.}~\bibnamefont
  {Tarnopolsky}}, \bibinfo {author} {\bibfnamefont {E.}~\bibnamefont {Khalaf}},
  \ and\ \bibinfo {author} {\bibfnamefont {A.}~\bibnamefont {Vishwanath}},\
  }\href@noop {} {\bibfield  {journal} {\bibinfo  {journal} {Physical Review
  Research}\ }\textbf {\bibinfo {volume} {2}},\ \bibinfo {pages} {023237}
  (\bibinfo {year} {2020})}\BibitemShut {NoStop}%
\bibitem [{\citenamefont {Dong}\ \emph
  {et~al.}(2023{\natexlab{b}})\citenamefont {Dong}, \citenamefont {Ledwith},
  \citenamefont {Khalaf}, \citenamefont {Lee},\ and\ \citenamefont
  {Vishwanath}}]{dong2023many}%
  \BibitemOpen
  \bibfield  {author} {\bibinfo {author} {\bibfnamefont {J.}~\bibnamefont
  {Dong}}, \bibinfo {author} {\bibfnamefont {P.~J.}\ \bibnamefont {Ledwith}},
  \bibinfo {author} {\bibfnamefont {E.}~\bibnamefont {Khalaf}}, \bibinfo
  {author} {\bibfnamefont {J.~Y.}\ \bibnamefont {Lee}}, \ and\ \bibinfo
  {author} {\bibfnamefont {A.}~\bibnamefont {Vishwanath}},\ }\href@noop {}
  {\bibfield  {journal} {\bibinfo  {journal} {Physical Review Research}\
  }\textbf {\bibinfo {volume} {5}},\ \bibinfo {pages} {023166} (\bibinfo {year}
  {2023}{\natexlab{b}})}\BibitemShut {NoStop}%
\end{thebibliography}
